\documentclass[twocolumn,aps,prd,amsmath,showpacs,nofootinbib]{revtex4}
\usepackage[10pt]{type1ec}  % use only 10pt fonts
\usepackage[T1]{fontenc}
\usepackage{CJKutf8}
\usepackage[overlap, CJK]{ruby}
\usepackage{CJKulem}
\newenvironment{SChinese}{%
  \CJKfamily{gbsn}%
  \CJKtilde
  \CJKnospace}{}
\usepackage[utf8]{inputenc}
\usepackage{graphicx}
\usepackage{bm}                      % bold math
\usepackage{mathptmx}                % math+times fonts
\usepackage{amssymb}  % \gtrsim, \geqslant, etc etc: see amsguide.ps
\usepackage{dcolumn}                 % Align table columns on decimal point
\usepackage{color}

\newcommand{\be}{\beta}

\newcommand{\De}{\Delta}
\newcommand{\ep}{\varepsilon}

\newcommand{\la}{\lambda}

\newcommand{\beq}{\begin{equation}}
\newcommand{\eeq}{\end{equation}}
\newcommand{\ba}{\begin{array}}
\newcommand{\ea}{\end{array}}
\newcommand{\bea}{\begin{eqnarray}}
\newcommand{\eea}{\end{eqnarray}}
\newcommand{\bi}{\begin{itemize}}  %\setlength{\itemsep}{0\parsep}}
\newcommand{\ei}{\end{itemize}}
\newcommand{\ben}{\begin{enumerate}} %\setlength{\itemsep}{0\parsep}}
\newcommand{\een}{\end{enumerate}}
\newcommand{\bc}{\begin{center}}
\newcommand{\ec}{\end{center}}

\newcommand\eqn[1]{(\ref{#1})}      % parentheses around the LaTex "ref" macro
\newcommand{\ee}[1]{\times 10^{#1}}

\newcommand{\fm}{{\rm fm}}

\newcommand{\MeV}{{\rm MeV}}
\newcommand{\GeV}{{\rm GeV}}

\newcommand{\ptrans}{p_{\rm trans}}
\newcommand{\ntrans}{n_{\rm trans}}
\newcommand{\etrans}{\varepsilon_{\rm trans}}

\newcommand{\Msolar}{{\rm M}_{\odot}}

\newcommand{\cQM}{{c^{\phantom{1}}_{\rm QM}}}

\newcommand{\cQMsq}{c^2_{\rm QM}}

\newcommand{\decrit}{\De\ep_{\rm crit}}
\newcommand{\Mmax}{M_{\rm max}}
\newcommand{\Rtyp}{R_{1.4}}

\def\bea{\begin{eqnarray}}
\def\eea{\end{eqnarray}}
\def\be{\begin{equation}}
\def\ee{\end{equation}}

\def\beq{\begin{equation}}
\def\eeq{\end{equation}}
\def\bar{\begin{array}[b]}
\def\barc{\begin{array}}
\def\bart{\begin{array}[t]}
\def\ear{\end{array}}

\begin{document}

\title{Constraining and applying a generic high-density equation of state}

\author{Mark G. Alford$^1$, G. F. Burgio$^2$, S. Han
(\begin{CJK}{UTF8}{}\begin{SChinese}韩 君\end{SChinese}\end{CJK})$^1$,
  G. Taranto$^{2,3}$, and D. Zappal\`a$^2$}

\affiliation{
$^1$Physics Department, Washington University, Saint Louis, Missouri
63130, USA}

\affiliation{
$^2$ INFN Sezione di Catania, Via Santa Sofia 64, 95123 Catania, Italy}

\affiliation{
$^3$Dipartimento di Fisica e Astronomia,
Universit\'a di Catania, Via Santa Sofia 64, 95123 Catania, Italy}

\date{1 Oct 2015} % hardwire date so arxiv can't change it

\begin{abstract}
We discuss the ``constant speed of sound'' (CSS) parametrization of the
equation of state of high-density matter and its application to the
field correlator method (FCM) model of quark matter.
We show how observational constraints on the maximum mass and typical radius
of neutron stars are expressed as constraints on the CSS parameters.
We find that the observation of a $2\,\Msolar$ star already severely
constrains the CSS parameters,  and is particularly difficult
to accommodate if the squared speed of sound in the high-density phase 
is assumed to be around $1/3$ or less.

We show that the FCM equation of state can be accurately represented
by the CSS parametrization, which assumes a sharp transition
to a high-density phase with density-independent speed of sound. 
We display the mapping between the
FCM and CSS parameters, and see that FCM only allows equations of state
in a restricted subspace of the CSS parameters.
\end{abstract}


\maketitle

%===============================================================================
\section{Introduction}
\label{sec:intro}

There are many models of matter at 
density significantly above nuclear saturation density,
each with their own parameters. In studying the equation of state (EoS)
of matter in this regime it is therefore
useful to have a general parametrization of the 
EoS which can be used as a generic language for relating
different models to each other and for expressing experimental constraints
in model-independent terms.
In this work we use the previously proposed
``constant speed of sound'' (CSS) parametrization 
\cite{Alford:2013aca,Zdunik:2012dj,Chamel:2012ea}
(for applications, see, e.g., \cite{Alvarez-Castillo:2014nua}).
We show how mass and radius observations can be expressed as
constraints on the CSS parameters.
Here we analyze a specific example, where the high-density matter
is quark matter described by a model based on the field correlator method
(Sec.~\ref{sec:FCM}), showing how its parameters can be mapped on
to the CSS parameter space, and how it is constrained by
currently available observations of neutron stars. 

The CSS parametrization is applicable to high-density equations of
state for which (a) there is a sharp interface between nuclear matter and
a high-density phase which we will call quark matter, even when
(as in Sec.~\ref{sec:CSS-constraints})
we do not make any assumptions about its physical nature; and (b) the speed of sound in the high-density
matter is pressure-independent for pressures ranging from the
first-order transition pressure up to the maximum central pressure
of neutron stars. One can then write
the high-density EoS in terms of three
parameters: the pressure $\ptrans$ of the transition, 
the discontinuity in energy density $\De\ep$ at the transition,
and  the speed of sound $\cQM$ in the high-density phase.
For a given nuclear matter EoS $\ep_{\rm NM}(p)$, the full CSS EoS is then
\beq
\ep(p) = \left\{\!
\begin{array}{ll}
\ep_{\rm NM}(p) & p<\ptrans \\
\ep_{\rm NM}(\ptrans)+\De\ep+c_{\rm QM}^{-2} (p-\ptrans) & p>\ptrans
\end{array}
\right.\ 
\label{eqn:CSS_EoS}
\eeq
The CSS form can be viewed as the lowest-order terms of a 
Taylor expansion of the
high-density EoS about the transition pressure.
Following Ref.~\cite{Alford:2013aca}, we express the three parameters
in dimensionless form, as $\ptrans/\etrans$, $\De\ep/\etrans$
(equal to $\la-1$ in the notation of Ref.~\cite{Haensel:1983})
and $\cQMsq$, where $\etrans \equiv \ep_{\rm NM}(\ptrans)$.

The assumption of a sharp interface will be valid if, for
example, there is a
first-order phase transition between nuclear and quark matter,
and the surface tension of the interface is high enough
to ensure that the transition occurs at a sharp interface (Maxwell
construction) not via a mixed phase (Gibbs construction). Given the uncertainties in the value
of the surface tension \cite{Alford:2001zr,Palhares:2010be,Pinto:2012aq},
this is a possible scenario. One can also formulate generic
equations of state that model 
interfaces that are smeared out by mixing or percolation 
\cite{Macher:2004vw,Masuda:2012ed, Alvarez-Castillo:2013spa}.

The assumption of a density-independent speed of sound is valid for a
large class of models of quark matter.
The CSS parametrization is an almost exact fit to some Nambu--Jona-Lasinio models \cite{Zdunik:2012dj,Agrawal:2010er,Bonanno:2011ch,Lastowiecki:2011hh}. The perturbative quark
matter EoS \cite{Kurkela:2010yk} also has roughly density-independent $\cQMsq$,
with a value around 0.2 to 0.3 (we use units where $\hbar=c=1$), above the transition from nuclear matter (see
Fig.~9 of Ref.~\cite{Kurkela:2009gj}).  In the quartic polynomial
parametrization \cite{Alford:2004pf}, varying the coefficient $a_{2}$ between
$\pm(150 {\rm MeV})^{2}$, and the coefficient $a_{4}$ between 0.6 and 1, and
keeping $\ntrans/n_0$ above 1.5 {($n_0\equiv 0.16\,\fm^{-3}$ is the
nuclear saturation density)}, one finds that $\cQMsq$ is always between 0.3
and 0.36. It is noticeable that models based on relativistic quarks
tend to have $\cQMsq\approx$ close to 1/3, which is the value for
systems with conformal symmetry, and it has been conjectured that 
there is a fundamental bound $\cQMsq<1/3$ \cite{Bedaque:2014sqa}, although some models violate that bound, e.g.~\cite{Kojo:2014rca,Benic:2014iaa} or \cite{Lastowiecki:2011hh} (parametrized in \cite{Zdunik:2012dj}).

In Sec.~\ref{sec:CSS-constraints} we 
show how the CSS parametrization is
constrained by observables such as the
maximum mass $\Mmax$, the radius of a maximum-mass star, and the radius $\Rtyp$
of a star of mass $1.4\,\Msolar$.
In Secs.~\ref{sec:BHF}\textendash\ref{sec:FCM} we describe a specific model,
based on a Brueckner-Hartree-Fock (BHF) calculation of the 
nuclear matter EoS and the field correlator method (FCM) for the quark
matter EoS. We show how the parameters of this model map on to part
of the CSS parameter space, and how the observational constraints apply to
the FCM model parameters.
Section \ref{sec:conclusions} gives our conclusions.

\section{Constraining the CSS parameters}
\label{sec:CSS-constraints}

\subsection{Topology of the mass-radius relation}

\begin{figure}[htb]
\includegraphics[width=0.9\hsize]{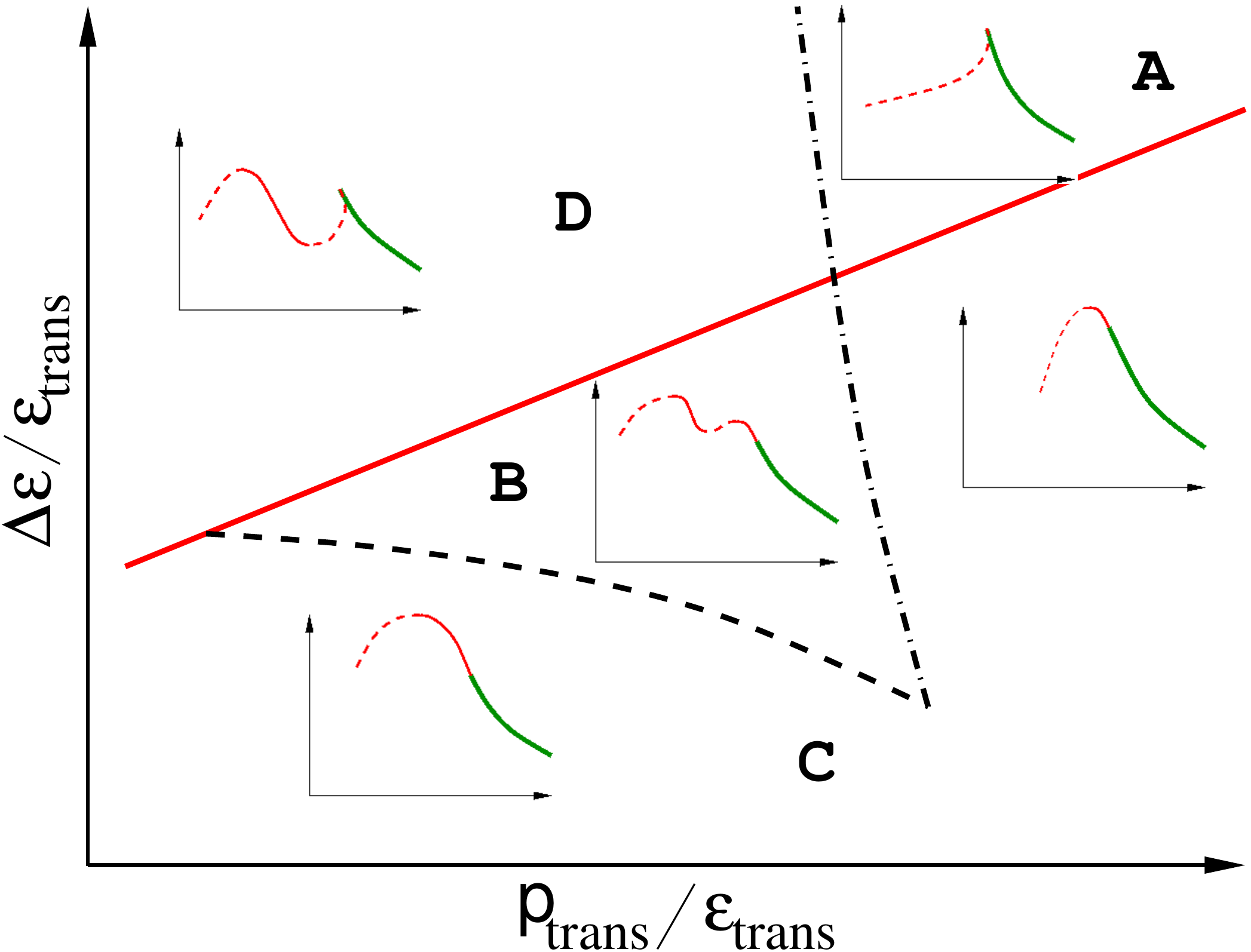}
\caption{(Color online). Schematic phase diagram (from \cite{Alford:2013aca})
for hybrid star branches in the
mass-radius relation of compact stars. We fix $\cQMsq$ and vary
$\ptrans/\etrans$ and $\De\ep/\etrans$. The four regions are
(A) no hybrid branch (``absent''); (B) both connected and
disconnected hybrid branches; (C) connected hybrid branch only; and (D) disconnected hybrid branch only.
}
\label{fig:phase-diag-schematic}
\end{figure}

We use the term ``hybrid star''
to refer to stars whose central pressure is above $\ptrans$, and so they
contain a core of the high-density phase. The part of the mass-radius relation
that arises from such stars is the hybrid branch.
In all models of nuclear/quark matter we find the same four topologies
of the mass-radius curve for compact stars: the hybrid branch
may be connected to the nuclear branch (C), or disconnected (D), or
both may be present (B) or neither (A). The occurrence of these
as a function of the CSS parameters $\ptrans/\etrans$ and $\De\ep/\etrans$
at fixed $\cQMsq$ is shown schematically in Fig.~\ref{fig:phase-diag-schematic}
(taken from Ref.~\cite{Alford:2013aca}).
The mass-radius curve in each region is depicted in inset plots,
in which the thick green line is the hadronic branch,
the thin solid red lines are stable hybrid stars, 
and the thin dashed red lines are unstable hybrid stars.

In the phase diagram 
the solid red line shows the threshold value $\decrit$ below which 
there is always a stable hybrid star branch connected to the neutron star 
branch. This critical value is given by 
\cite{Seidov:1971,Haensel:1983,Lindblom:1998dp}
\be
\frac{\decrit}{\etrans} = \frac{1}{2} +  \frac{3}{2}   \frac{\ptrans}{\etrans}  
\label{e:crit}
\ee
and was obtained by performing an expansion in powers of the size of the 
core of high-density phase. Equation~\eqn{e:crit} is an analytic result,
independent of $\cQMsq$ and the nuclear matter EoS.
The dashed and dot-dashed black lines mark the border of regions where the disconnected hybrid star branch exists. The position of these lines depends on the value of $\cQMsq$ and
(weakly) on the accompanying nuclear matter EoS \cite{Alford:2013aca}.

Once a nuclear matter EoS has been chosen, any high-density EoS that
is well approximated by the CSS parametrization can be summarized by
giving the values of the three CSS parameters, 
corresponding to a point in the phase diagram.
We then know what sort
of hybrid branches will be present. 

\subsection{Maximum mass of hybrid stars}
\label{sec:maxmass}

\begin{figure*}[htb]
\parbox{0.5\hsize}{
\includegraphics[width=\hsize]{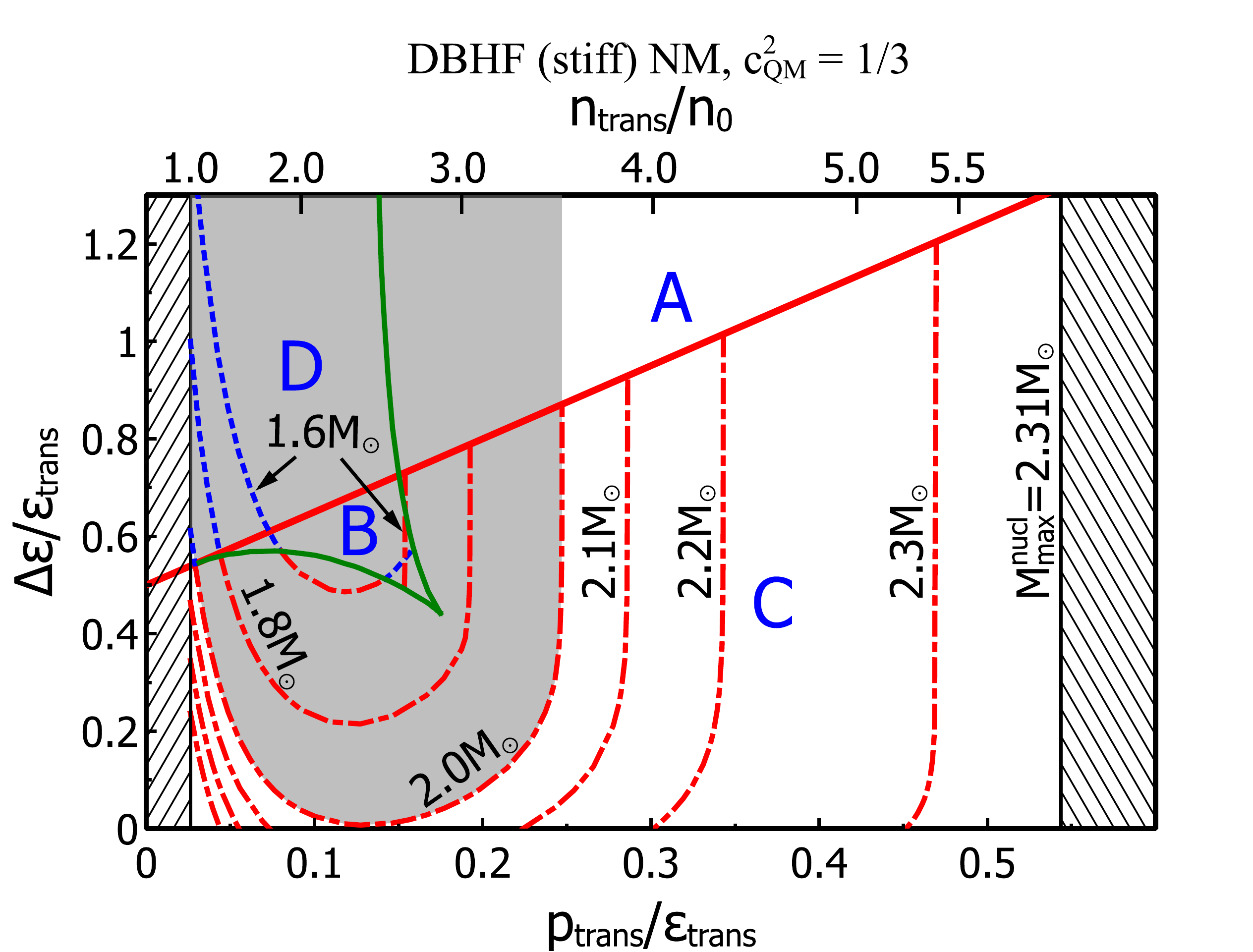}\\[2ex]
}\parbox{0.5\hsize}{
\includegraphics[width=\hsize]{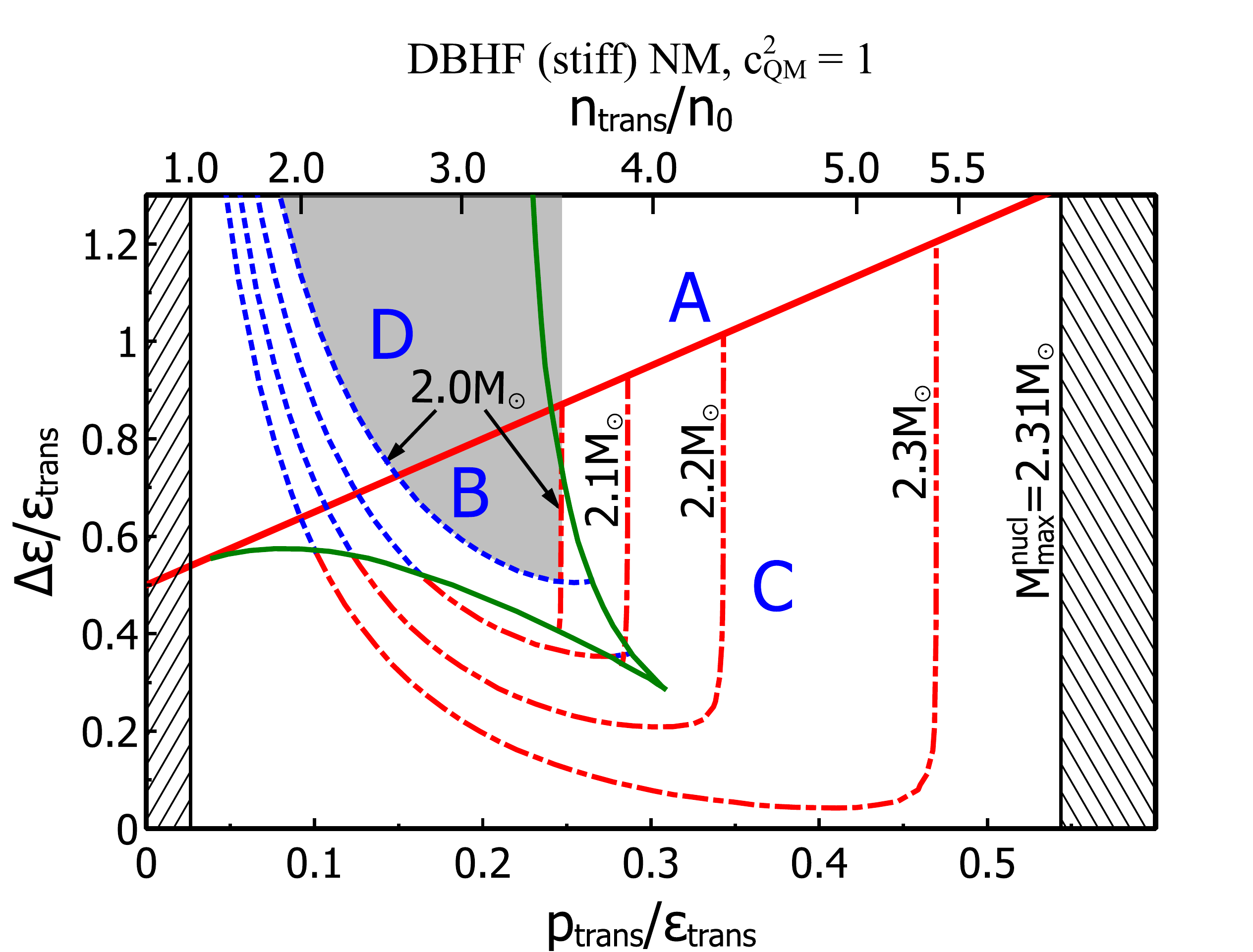}\\[2ex]
}\\[2ex]
\parbox{0.5\hsize}{
\includegraphics[width=\hsize]{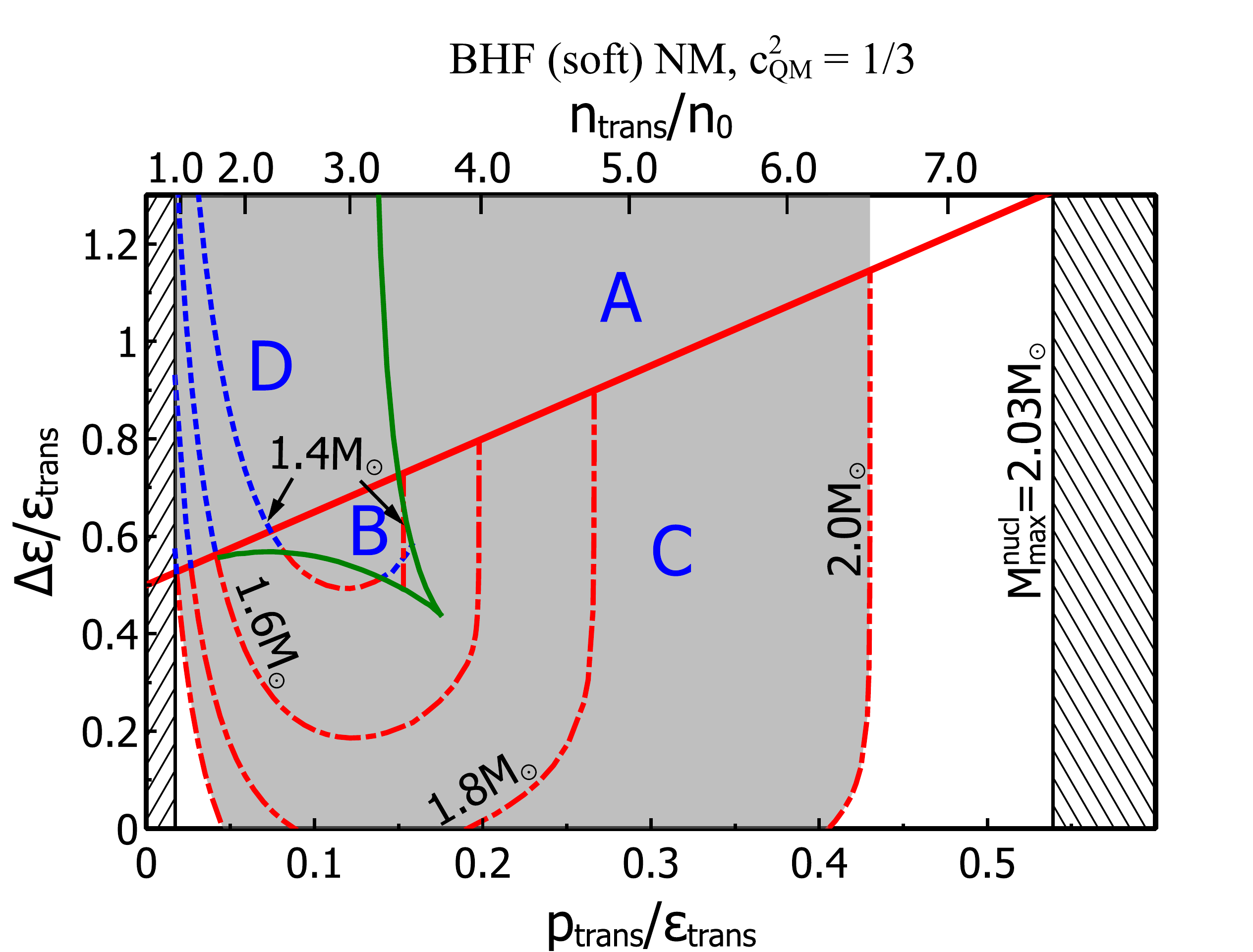}\\[2ex]
}\parbox{0.5\hsize}{
\includegraphics[width=\hsize]{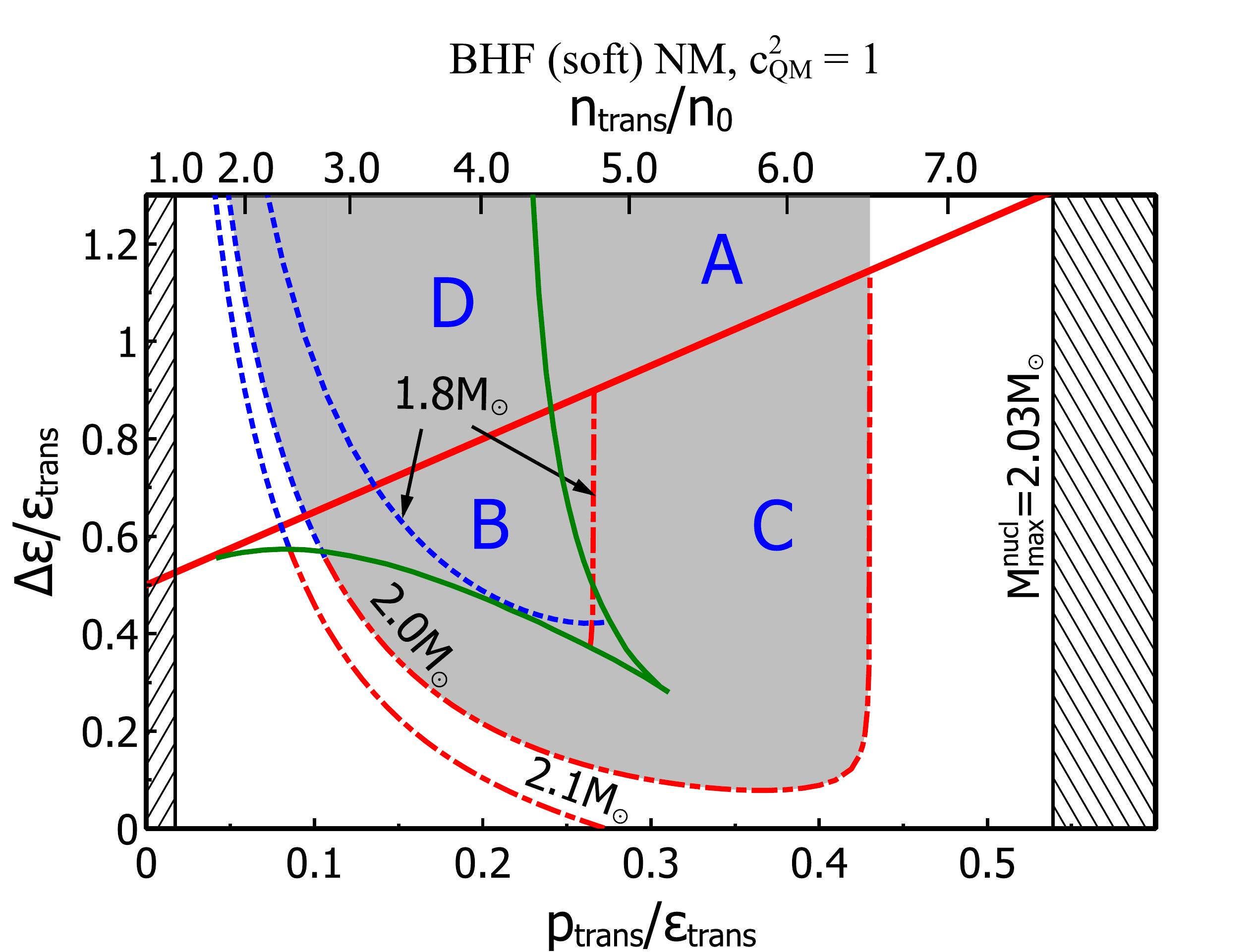}\\[2ex]
}
\caption{(Color online). Contour plots showing the maximum hybrid star mass as a function of the
CSS parameters of the high-density EoS. Each panel shows the dependence
on the CSS parameters $\ptrans/\etrans$ and $\De\ep/\etrans$. The left
plots are for  $\cQMsq=1/3$, and the right plots are for $\cQMsq=1$.
The top row is for a DHBF (stiff) nuclear matter EoS, and the bottom row
is for a BHF (soft) nuclear matter EoS.
The grey shaded region 
is excluded by the measurement of a $2\,\Msolar$ star.
The hatched band at low density (where $\ntrans<n_0$) is
excluded because bulk nuclear matter would be metastable. The
hatched band at high density is excluded because the transition pressure
is above the central pressure of the heaviest stable hadronic star.
}
\label{fig:CSS-max-mass}
\end{figure*}

\begin{table}[htb]
\begin{center}
\begin{tabular}{c c@{\quad} c}
\hline
Property & BHF, \, Av${}_{18}$  & DBHF, \\
         & + UVIX~TBF           & Bonn~A \\
\hline
Saturation baryon density $n_0 \rm (fm^{-3})$   & 0.16 & 0.18  \\
Binding energy/baryon $E/A$ (MeV)     & -15.98  &  -16.15  \\
Compressibility $K_0$ (MeV)           & 212.4 &  230  \\
Symmetry energy $S_0 $ (MeV)          & 31.9  &  34.4  \\
$L = 3 n_0  \, [ dS_0/dn ]_{n_0}$ (MeV)    & 52.9  &  69.4 \\
Maximum mass of star ($\Msolar$)          & 2.03 & 2.31 \\
Radius of the heaviest star (km)  & 9.92 &11.26 \\
Radius of $M=1.4\,\Msolar$ star (km)  & 11.77 &13.41 \\
\hline
\end{tabular}
\end{center}
\caption{(Color online). Calculated properties of symmetric nuclear matter for the BHF and Dirac-Brueckner-Hatree-Fock (DBHF)
nuclear equations of state used here. BHF is softer, and DBHF is stiffer
(see Sec.~\ref{sec:BHF})}
\label{tab:EoS}
\end{table}  

In Fig.~\ref{fig:CSS-max-mass} we show how mass measurements of
neutron stars can be expressed as constraints on the CSS parameters.
Each panel shows dependence on $\ptrans/\etrans$ and $\De\ep/\etrans$
for fixed $\cQMsq$, as in Fig.~\ref{fig:phase-diag-schematic}.
The region in which the transition to quark matter would occur below
nuclear saturation density ($\ntrans<n_0$) is excluded (hatched band
at the left end) because
in that region bulk nuclear matter would be metastable.
There is also an upper limit on the transition pressure, which is
the central pressure of the heaviest stable nuclear matter star.
This depends on the hadronic EoS that had been assumed.

The contours show the maximum mass of a hybrid star as a function of
the EoS parameters.
The region inside the $M=2\,\Msolar$ contour corresponds to EoSes for which
the maximum mass is less than $2\,\Msolar$ so it is shaded to signify that this
region of parameter space for the high-density EoS is excluded by the
observation of a star with mass $2\,\Msolar$ \cite{Antoniadis:2013en}.
For high-density EoSs with $\cQMsq=1$ (right-hand plots), this
region is not too large, and leaves a good range of transition
pressures and energy density discontinuities that are compatible
with the observation.
However, for high-density matter with $\cQMsq=1/3$ (left-hand plots), 
which is the typical value in many models
(See Sec.~\ref{sec:intro}), the  $\Mmax>2\,\Msolar$ constraint eliminates
a large region of the CSS parameter space
\cite{Alford:2013aca,Bedaque:2014sqa}. 
We discuss this in more detail below.

The upper plots in Fig.~\ref{fig:CSS-max-mass}
are for a stiffer nuclear matter EoS, DBHF\cite{GrossBoelting:1998jg}, and
the lower plots are for a softer nuclear matter EoS, BHF \cite{Taranto:2013gya} (see Sec.~\ref{sec:BHF}).
Properties of these nuclear matter EoSs are given in Table \ref{tab:EoS}.
As one would expect, the stiffer EoS gives rise to heavier
(and larger) stars,
and therefore allows a wider range of CSS parameters to be compatible
with the $2\,\Msolar$ measurement.

In Fig.~\ref{fig:CSS-max-mass} the dot-dashed (red)
contours are for hybrid stars on a connected branch, while the
dashed (blue) contours are for disconnected branches. As discussed
in Ref.~\cite{Alford:2013aca}, when crossing the near-horizontal boundary 
from region C to B the connected hybrid branch splits into a smaller
connected branch and a disconnected branch, so the maximum mass
of the connected branch smoothly becomes the maximum mass of the 
disconnected branch. Therefore the red contour in the C region
smoothly becomes a blue contour in the B and D regions.
When crossing the near-vertical  boundary from region C to B a
new disconnected branch forms, so the connected branch (red dot-dashed) contour 
crosses this boundary smoothly.

In each panel of 
Fig.~\ref{fig:CSS-max-mass}, the physically relevant allowed region is
the white unshaded region.
The grey shaded region is excluded by the existence of a $2\,\Msolar$ 
star. We see that increasing the stiffness of the hadronic EoS or
of the quark matter EoS (by increasing $\cQMsq$) shrinks the excluded
region.

For both the hadronic EoSs that we study, the CSS parameters are
significantly constrained. From the two left panels of
Fig.~\ref{fig:CSS-max-mass} one can see that if,
as predicted by many models, $\cQMsq\lesssim 1/3$,
then we are limited to two regions of parameter space,
corresponding to a lowpressure transition or a high pressure transition.
In the low-transition-pressure region the transition occurs
at a fairly low density $\ntrans \lesssim 2\,n_0$, and a connected hybrid
branch is possible. In the high-transition-pressure region
the connected branch
(red dot-dashed) contours are, except at very low $\De\ep$, almost vertical,
corresponding to EoSs that give rise to a very small connected hybrid branch
which exists in a very small range of central pressures $p_{\rm cent}$ just above $\ptrans$.
The maximum mass on this branch
is therefore very close to the mass of the purely hadronic matter
star  with  $p_{\rm cent}=\ptrans$. 
The mass of such a purely hadronic star is naturally independent
of parameters that only affect the quark matter EoS, such as 
$\De\ep$ and $\cQMsq$, so the contour is vertical. These hybrid stars
have a tiny core of the high-density phase and cover a tiny range of masses,
of order $10^{-3}\,\Msolar$ or less, and so would be very rare.

Disconnected hybrid branches are of special interest, because they
give a characteristic signature in mass-radius measurements.
For both the hadronic EoSs that we study,
the region B and D, where disconnected hybrid star branches
can occur, are excluded for $\cQMsq\leqslant 1/3$. Even for
larger $\cQMsq$ disconnected branches 
only arise if the nuclear matter EoS is
sufficiently stiff. It is interesting to note that using an
extremely stiff hadronic matter EoS such as DD2-EV \cite{Benic:2014jia}
can further shrink the region that is excluded by the 
$\Mmax>2\,\Msolar$ constraint, allowing disconnected branches of hybrid stars
to occur.

\begin{figure*}[htb]
\parbox{0.5\hsize}{
\includegraphics[width=\hsize]{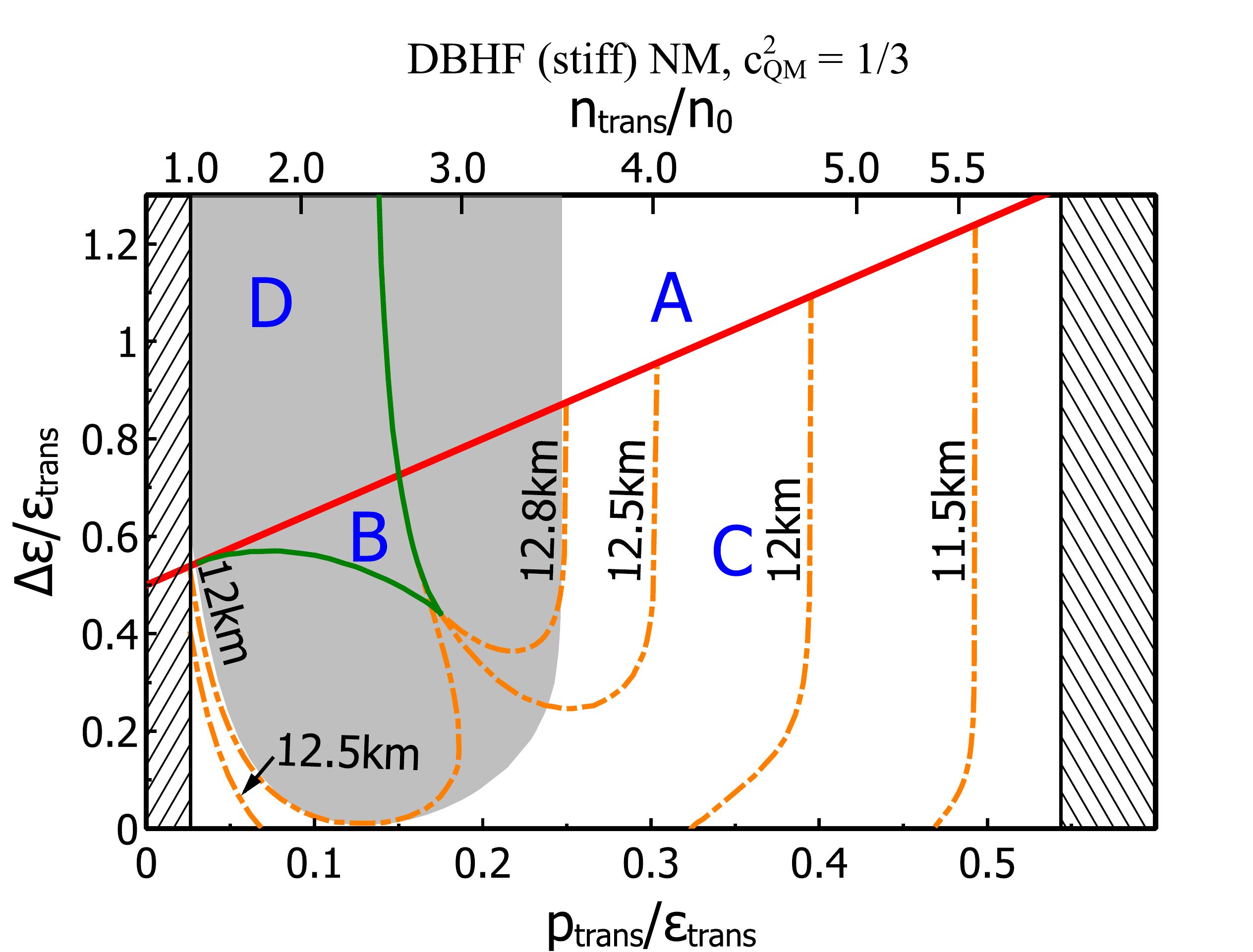}\\[2ex]
}\parbox{0.5\hsize}{
\includegraphics[width=\hsize]{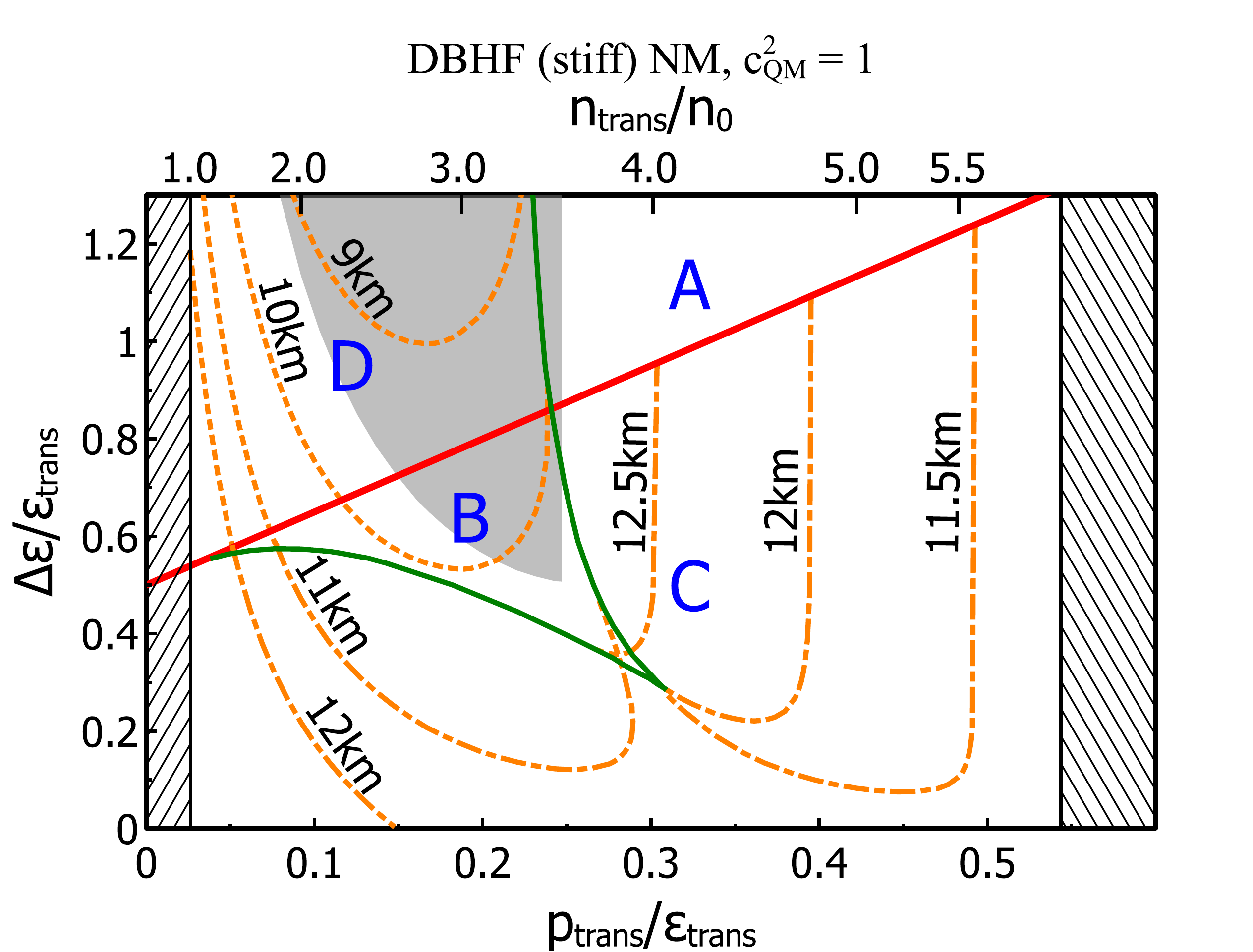}\\[2ex]
}\\[2ex]
\parbox{0.5\hsize}{
\includegraphics[width=\hsize]{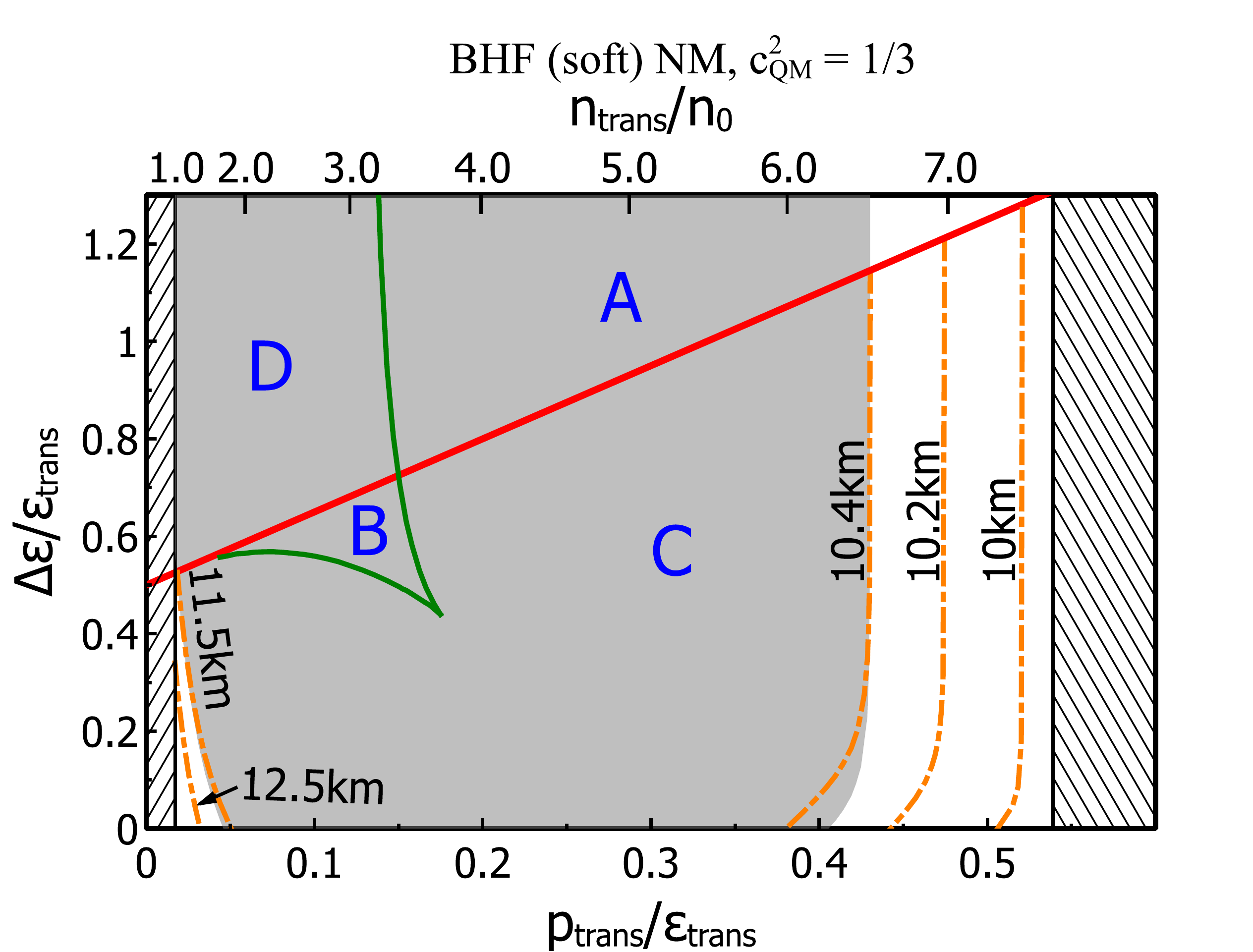}\\[2ex]
}\parbox{0.5\hsize}{
\includegraphics[width=\hsize]{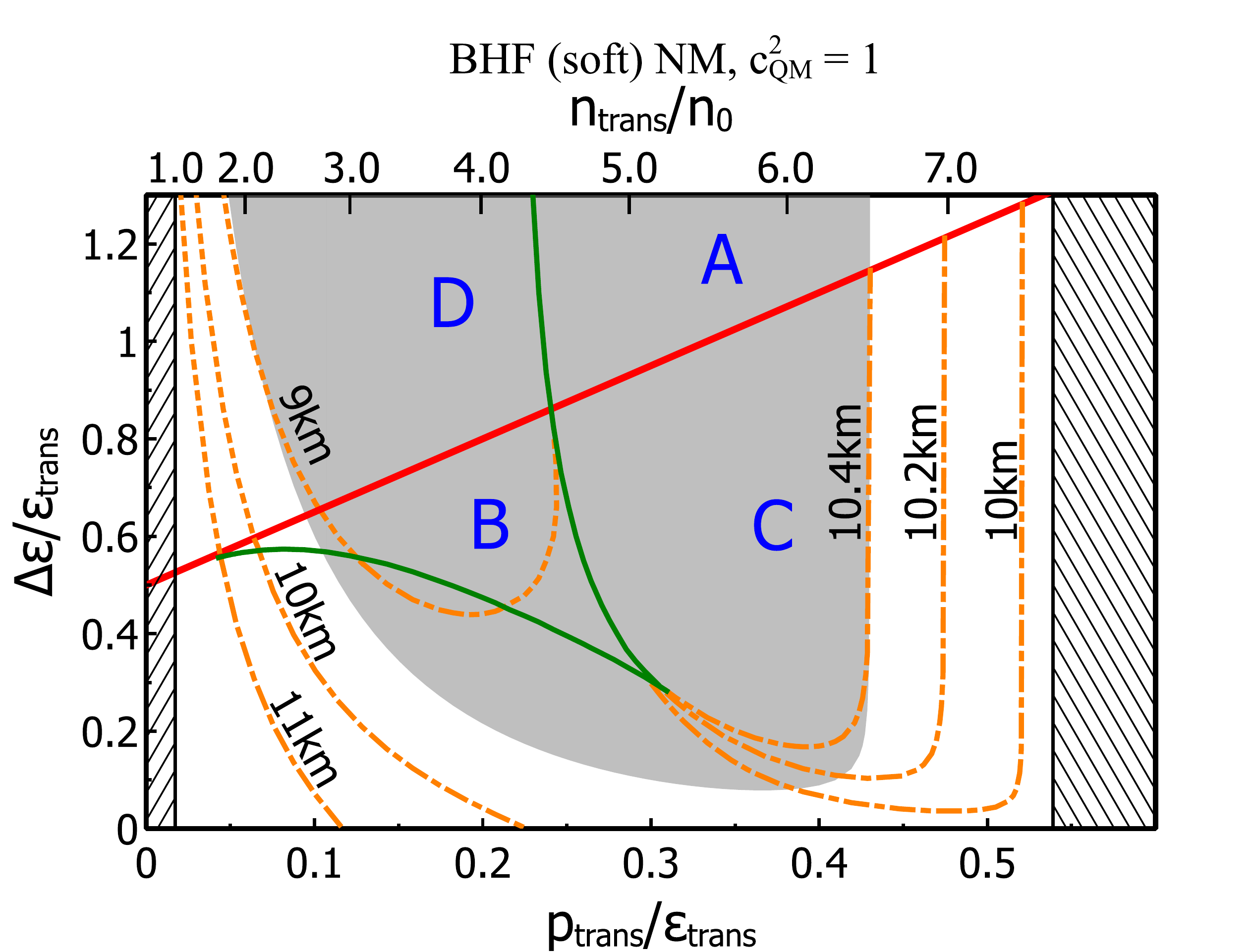}\\[2ex]
}
\caption{(Color online). Contour plots showing the radius of the maximum-mass star as a function of the
CSS parameters. Dashed lines are for the case where this star is on the
disconnected branch; for dot-dashed lines it is on the connected branch.
The grey shaded region 
is excluded by the measurement of a $2\,\Msolar$ star.
The hatched band at low density (where $\ntrans<n_0$) is
excluded because bulk nuclear matter would be metastable. The
hatched band at high density is excluded because the transition pressure
is above the central pressure of the heaviest stable hadronic star.
For a magnified version of the low-transition-pressure region for $\cQMsq=1/3$,
see Fig.~\ref{fig:CSS-radius-zoom}.
}
\label{fig:CSS-radius-max-mass}
\end{figure*}

\begin{figure*}[htb]
\parbox{0.5\hsize}{
%\centerline{DBHF (stiff) NM,\quad $\cQMsq=1/3$}
\includegraphics[width=\hsize]{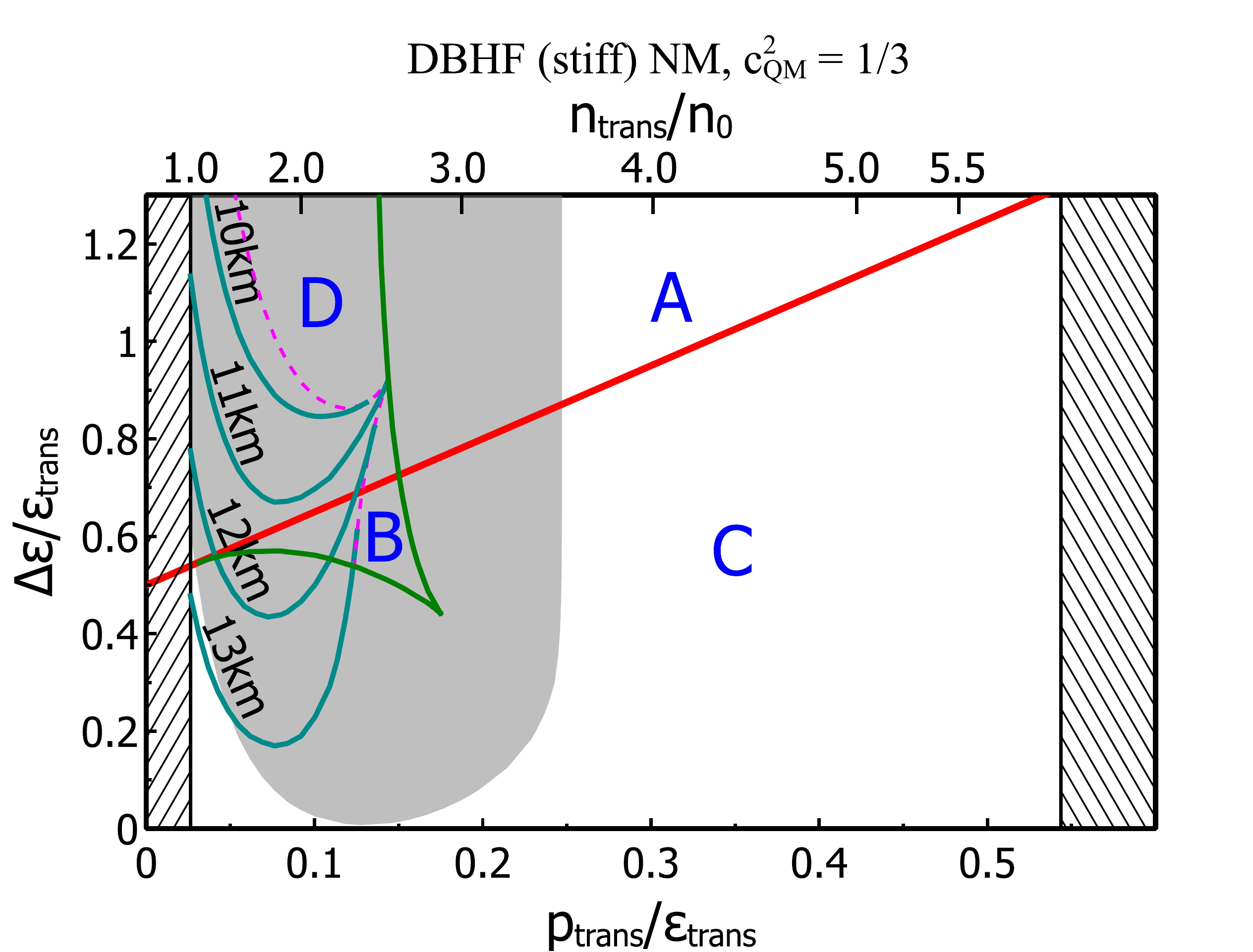}
}\parbox{0.5\hsize}{
%\centerline{DBHF (stiff) NM,\quad $\cQMsq=1$}
\includegraphics[width=\hsize]{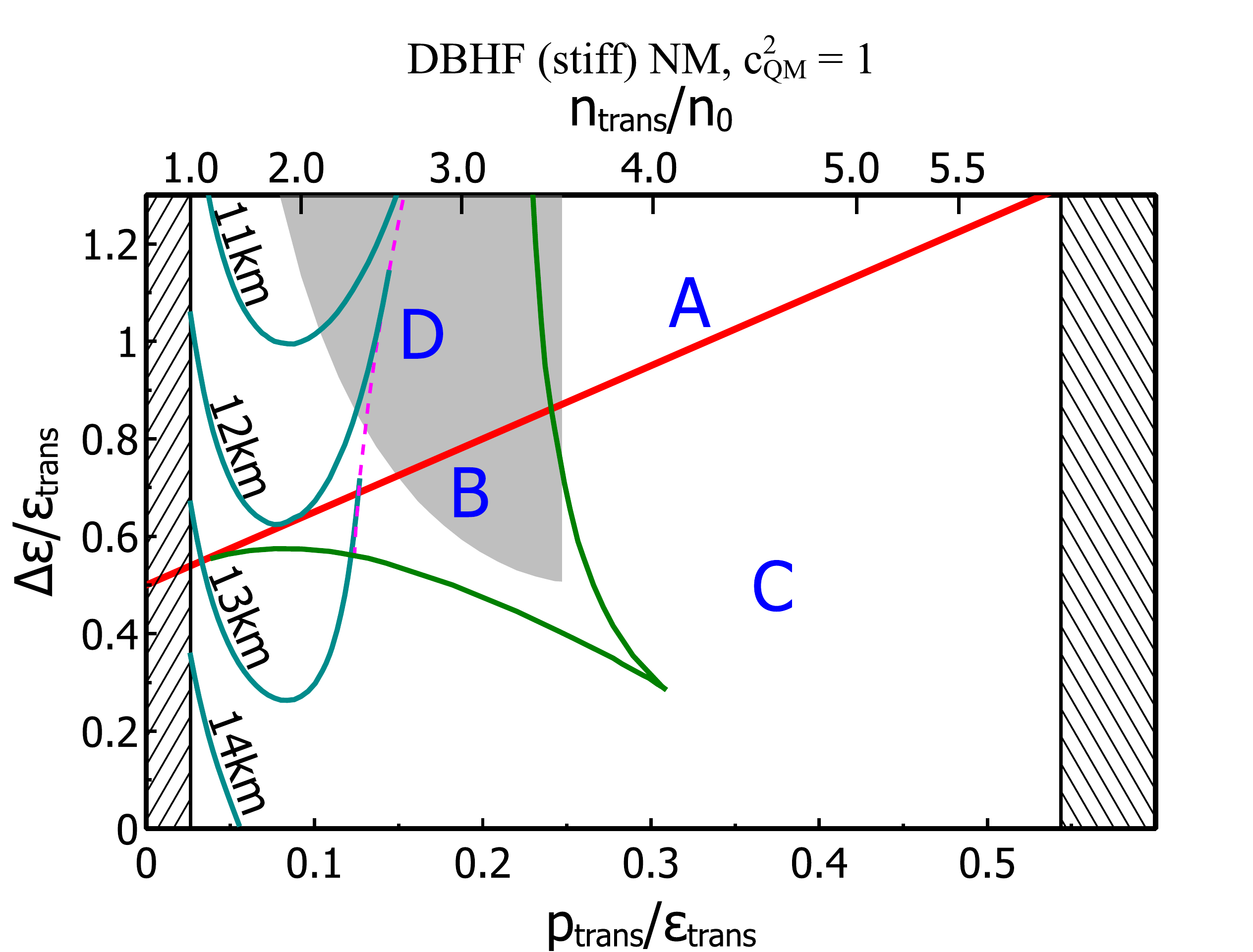}
}\\[2ex]
\parbox{0.5\hsize}{
%\centerline{BHF (soft) NM,\quad $\cQMsq=1/3$}
\includegraphics[width=\hsize]{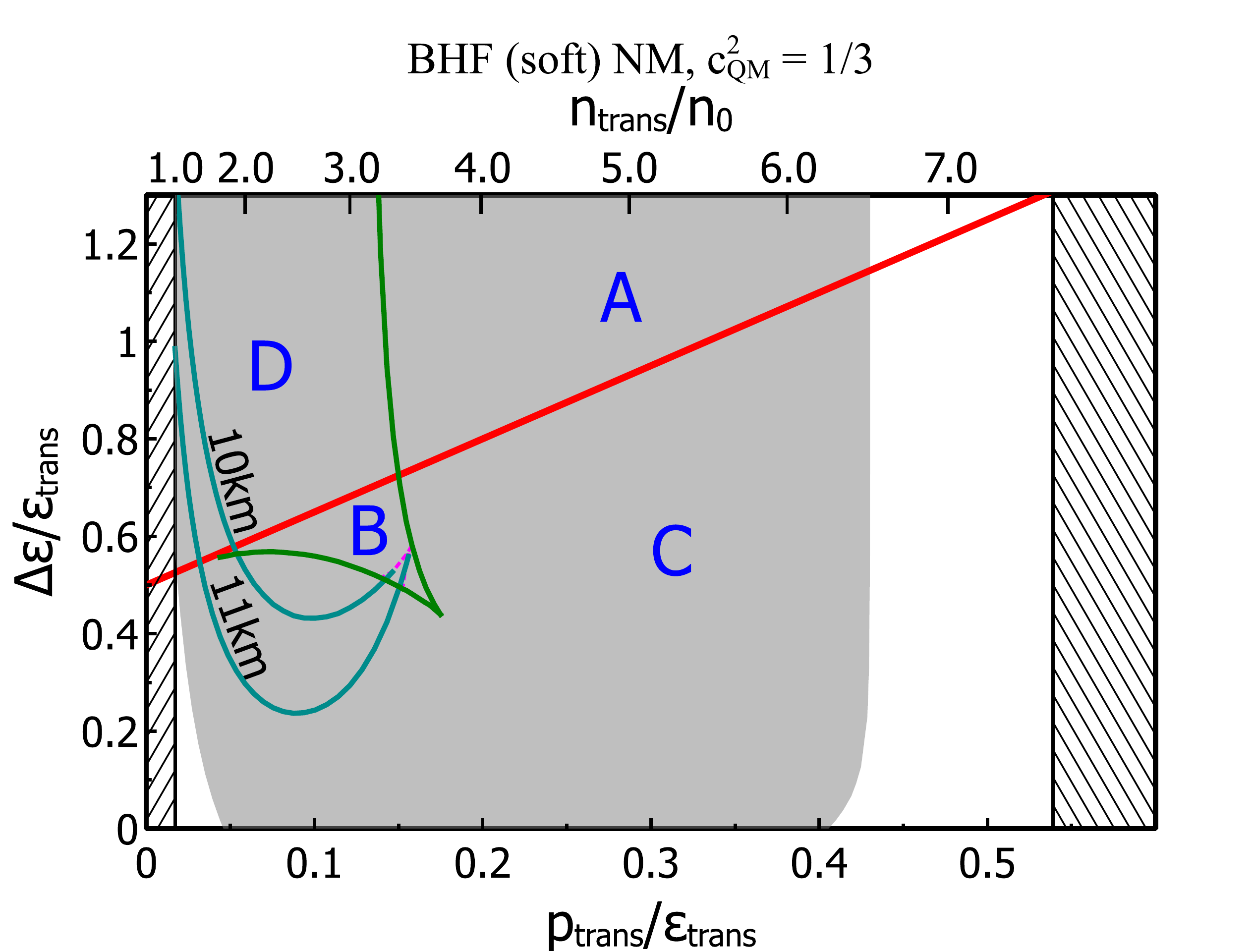}
}\parbox{0.5\hsize}{
%\centerline{BHF (soft) NM,\quad $\cQMsq=1$}
\includegraphics[width=\hsize]{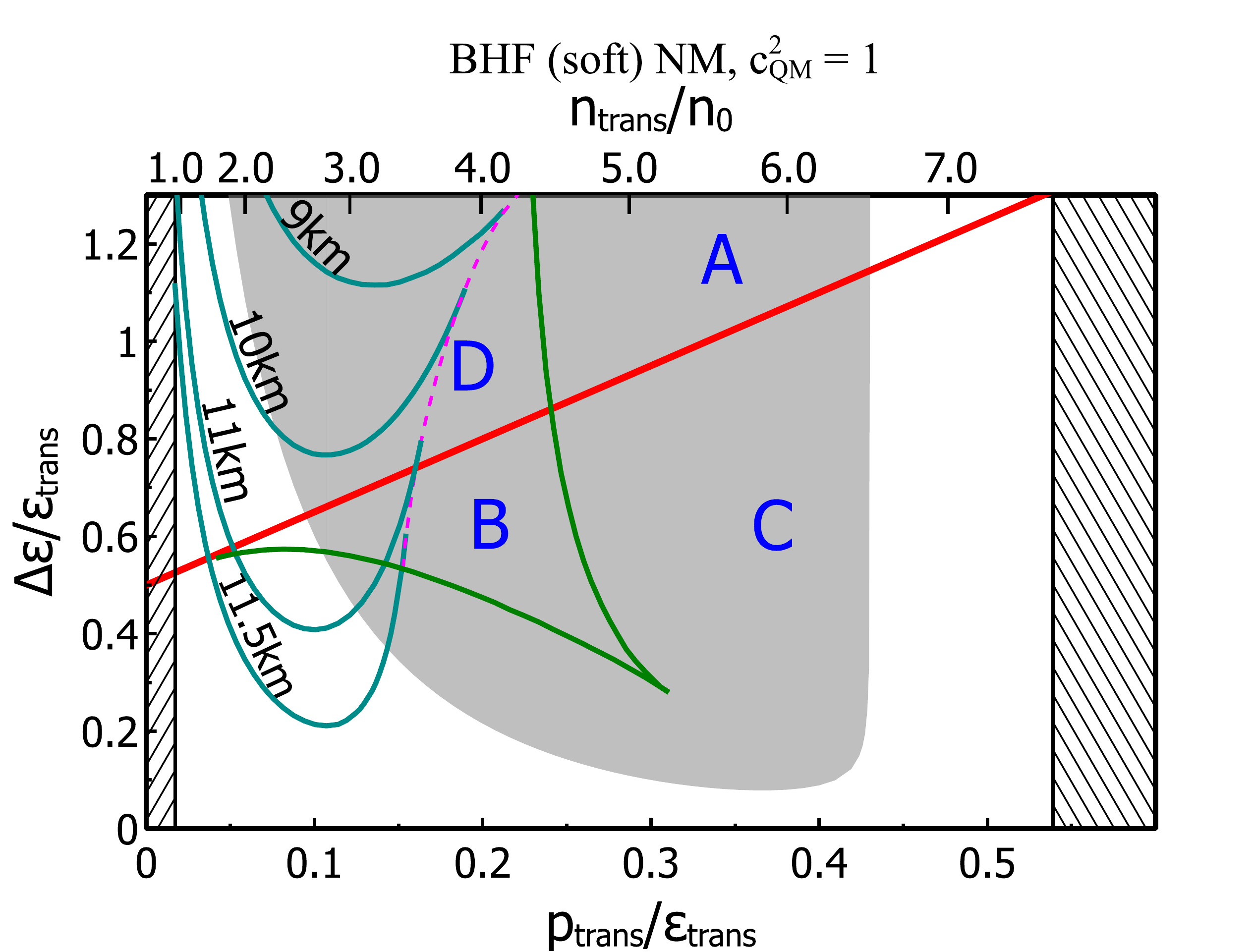}
}
\caption{(Color online). Contour plots similar to Fig.~\ref{fig:CSS-radius-max-mass}
showing the radius of a hybrid star of mass $M=1.4\,\Msolar$
as a function of the CSS parameters. Such stars only exist in a limited
region of the space of EoSs [delimited by dashed (magenta) lines].
The grey shaded region is excluded by the 
observational constraint $\Mmax>2\,\Msolar$.
For a magnified version of the low-transition-pressure region for $\cQMsq=1/3$,
see Fig.~\ref{fig:CSS-radius-zoom}.
}
\label{fig:CSS-radius}
\end{figure*}

\subsection{Minimum radius of hybrid stars}

In Fig.~\ref{fig:CSS-radius-max-mass} we show contour plots of the 
radius of the maximum-mass star (on either a connected or disconnected hybrid branch) as a function of the CSS quark matter
EoS parameters. Since the smallest hybrid star is typically the heaviest
one, this allows us to infer the smallest radius that arises from
a given EoS.

The layout is as in  Fig.~\ref{fig:CSS-max-mass}:
each panel shows dependence on $\ptrans/\etrans$ and $\De\ep/\etrans$
for fixed $\cQMsq$; the plots on the left are for $\cQMsq=1/3$ 
and the plots on the right are for $\cQMsq=1$; the plots on the top
are for the stiffer DBHF nuclear matter EoS, while
the lower plots are for the softer BHF nuclear matter EoS.
As in Fig.~\ref{fig:CSS-max-mass},
the region that is eliminated by the observation of a $2\,\Msolar$
star is shaded in grey.

The smallest stars, with radii as small as 9\,km, occur when
the high-density phase has the largest possible speed of sound
$\cQMsq=1$. They are disconnected branch stars arising from EoSs having
a low transition pressure ($\ntrans \lesssim 2\,n_0$) 
with a fairly large energy density discontinuity 
($\De \ep/\etrans \gtrsim 1$).

As in Fig.~\ref{fig:CSS-max-mass}, the contours in the
high-transition-pressure region are almost vertical because the hybrid branch
is then a very short extension to the nuclear mass-radius relation, and its
radius is close to that of the heaviest purely hadronic star, which is
independent of $\De\ep/\etrans$ and $\cQMsq$. The radius of the hybrid stars
decreases with $\ptrans$ in this region, because
the radius of hadronic stars decreases with central pressure.

For $\cQMsq=1/3$, the allowed low-transition-pressure region is
disconnected from the high-transition-pressure region and is
so small that it is hard to see on this plot. By magnifying it (left-hand
plots of Fig.~\ref{fig:CSS-radius-zoom}) we see that in this region
the radius contours closely track the border of the allowed region
(the $M_{\rm max}=2\,\Msolar$ line)
so we can say that the radius must be greater than 11.5\,km almost
independent of the transition pressure and hadronic EoS. 
For a stiff hadronic EoS this
minimum is raised to 11.7\,km. These values are comparable to
the minimum
radius of about 11.8\,km found in Ref.~\cite{Bedaque:2014sqa}, which
explored a larger set of hadronic EoSs but did not
explore the full CSS parameter space for the high-density EoS.
If a star with radius smaller than this minimum value were to be observed,
we would have to conclude that either the transition
occurs outside the low-density region or that $\cQMsq$ is greater than 1/3.
In the magnified figure we also show how the excluded region would
grow if a $2.1\,\Msolar$ star were to be observed (long-dashed line
for connected branch stars and short-dashed line for disconnected branch stars).
This would increase the minimum radius to about 12.1\,km for the soft hadronic EoS and 12.2\,km for the stiff hadronic EoS.

\subsection{Typical radius of hybrid stars}

In Fig.~\ref{fig:CSS-radius} we show contours (the U-shaped lines) of
typical radius of a hybrid star, defined as
$R_{1.4}$, the radius of a star of mass $1.4\,\Msolar$,
as a function of the CSS quark matter EoS parameters. 
The contours only fill the part of the CSS parameter space
where there are hybrid stars with that mass.
The dashed (magenta) lines delimit that region
which extends only up to moderate transition pressure.

The overall behavior is that, at fixed $\De\ep/\etrans$, the typical radius
is large when the transition density is at its lowest. As the transition
density rises the radius of a $1.4\,\Msolar$ star decreases at first, but then 
increases again. This is related to the
previously noted fact \cite{Yudin:2014}
that when one fixes the speed of sound of quark
matter and increases the bag constant 
(which increases $\ptrans/\etrans$ and also varies
$\De\ep/\etrans$ in a correlated way)
the resultant family of mass-radius curves
all pass through the same small region
in the $M$-$R$ plane: the $M(R)$ curves ``rotate'' counterclockwise around
this hub (see Fig.\,2 of Ref.~\cite{Yudin:2014}).
In our case we are varying $\ptrans/\etrans$ at fixed $\De\ep/\etrans$,
so the hub itself also moves. At low transition density the hub is
below $1.4\,\Msolar$, so $R_{1.4}$ decreases with $\ptrans/\etrans$.
At high transition density the hub is at a mass above $1.4\,\Msolar$ 
so $R_{1.4}$ will increase with $\ptrans/\etrans$.

The smallest stars occur for $\cQMsq=1$ (right-hand plots), 
where $R_{1.4}\gtrsim 9.5\,{\rm km}$
at large values of the energy density discontinuity, and the radius
rises as the discontinuity is decreased. This is consistent
with the absolute lower bound of about 
8.5\,km \cite{Lattimer:2012nd} for the
maximally compact $\cQMsq=1$ star obeying  $\Mmax>2\,\Msolar$.

\begin{figure*}[htb]
\parbox{0.5\hsize}{
\includegraphics[width=\hsize]{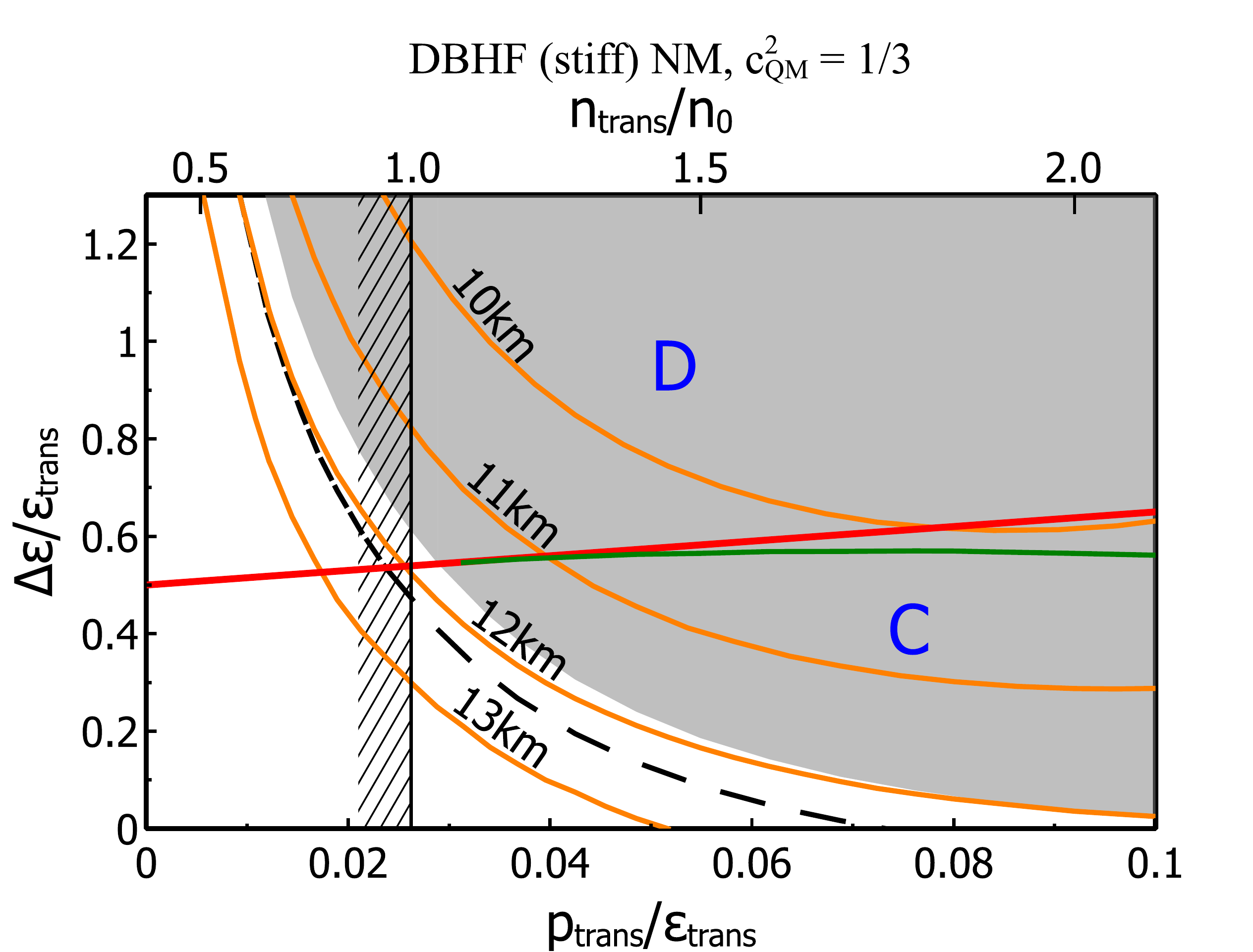}
}\parbox{0.5\hsize}{
\includegraphics[width=\hsize]{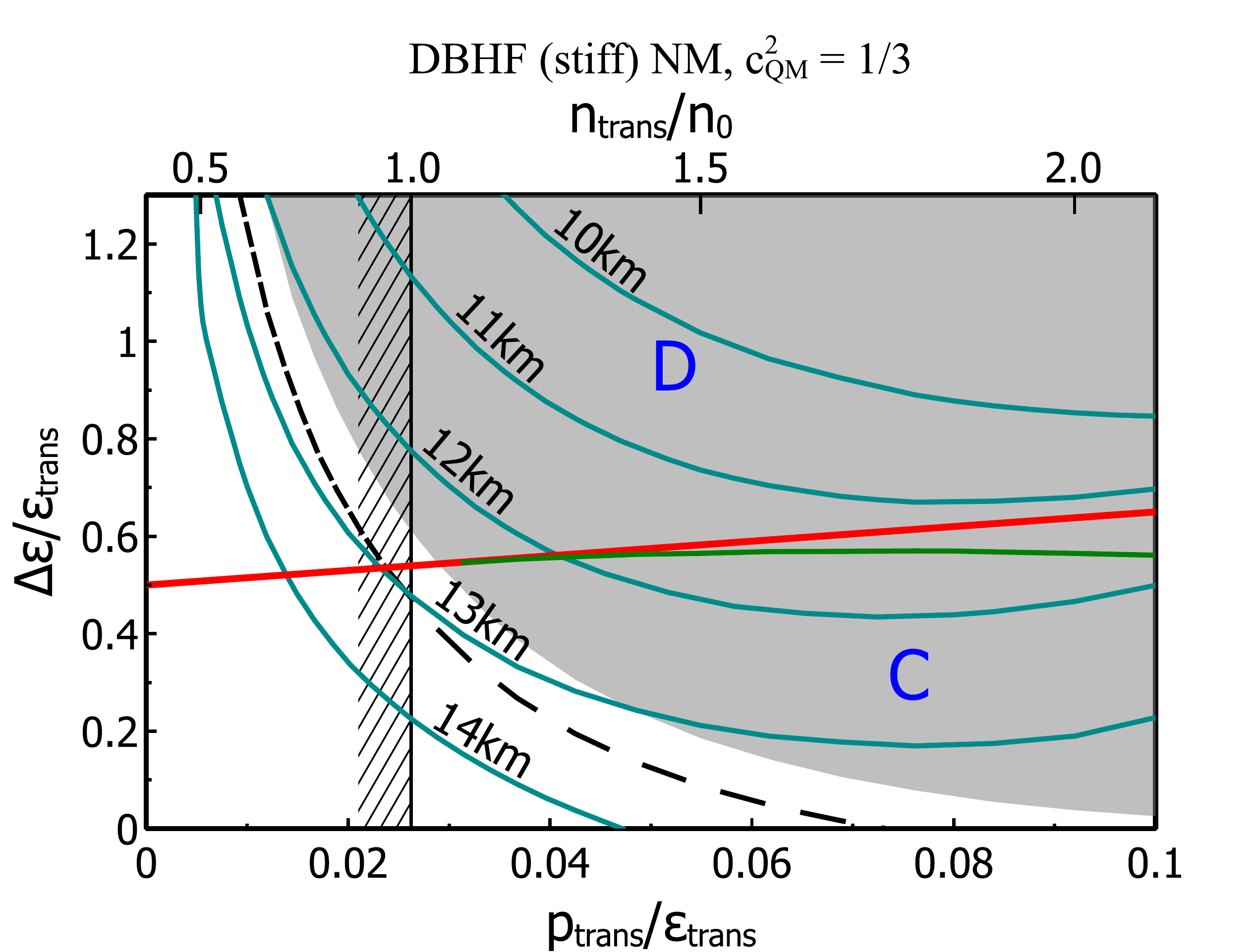}
}\\[2ex]
\parbox{0.5\hsize}{\includegraphics[width=\hsize]{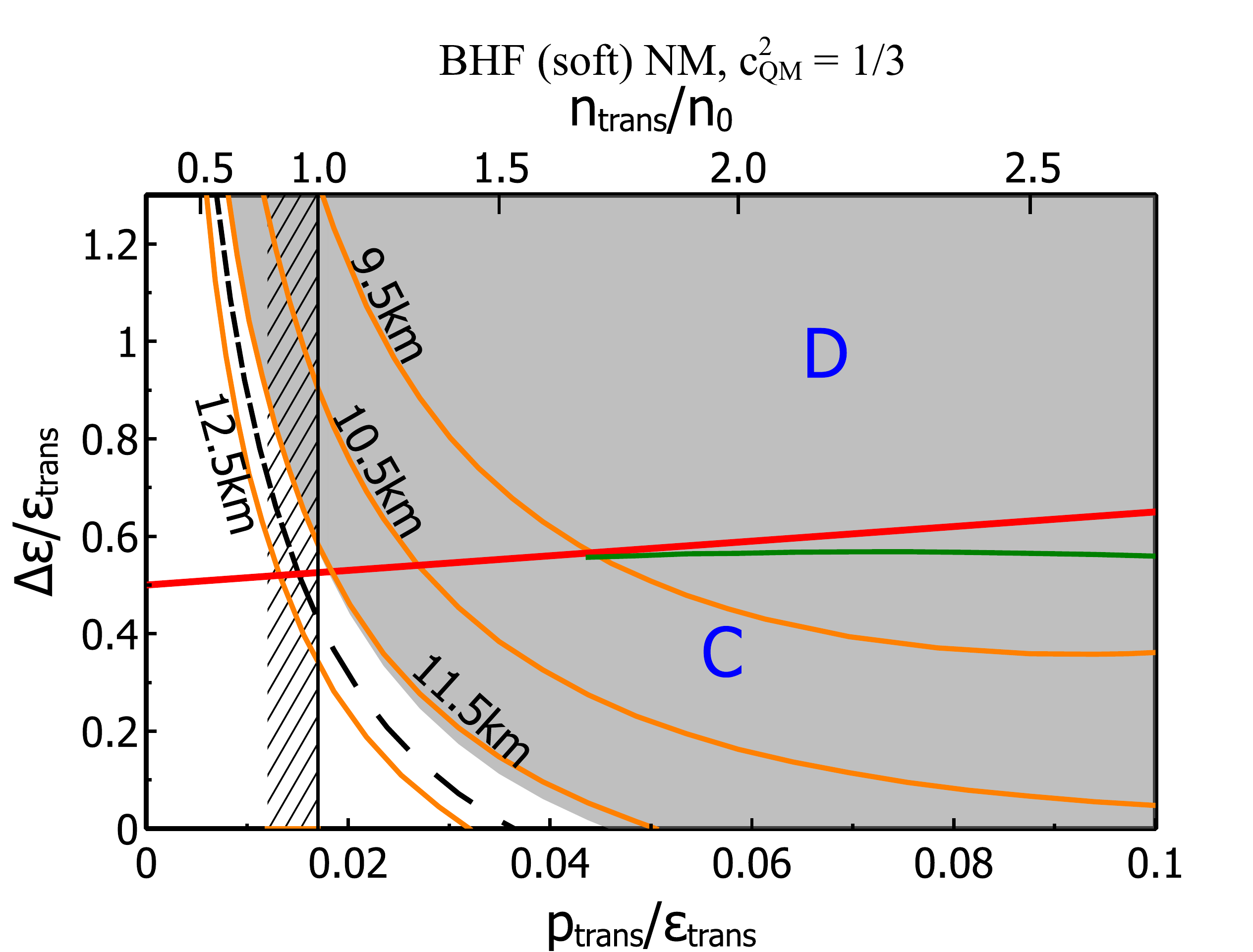}
}\parbox{0.5\hsize}{
\includegraphics[width=\hsize]{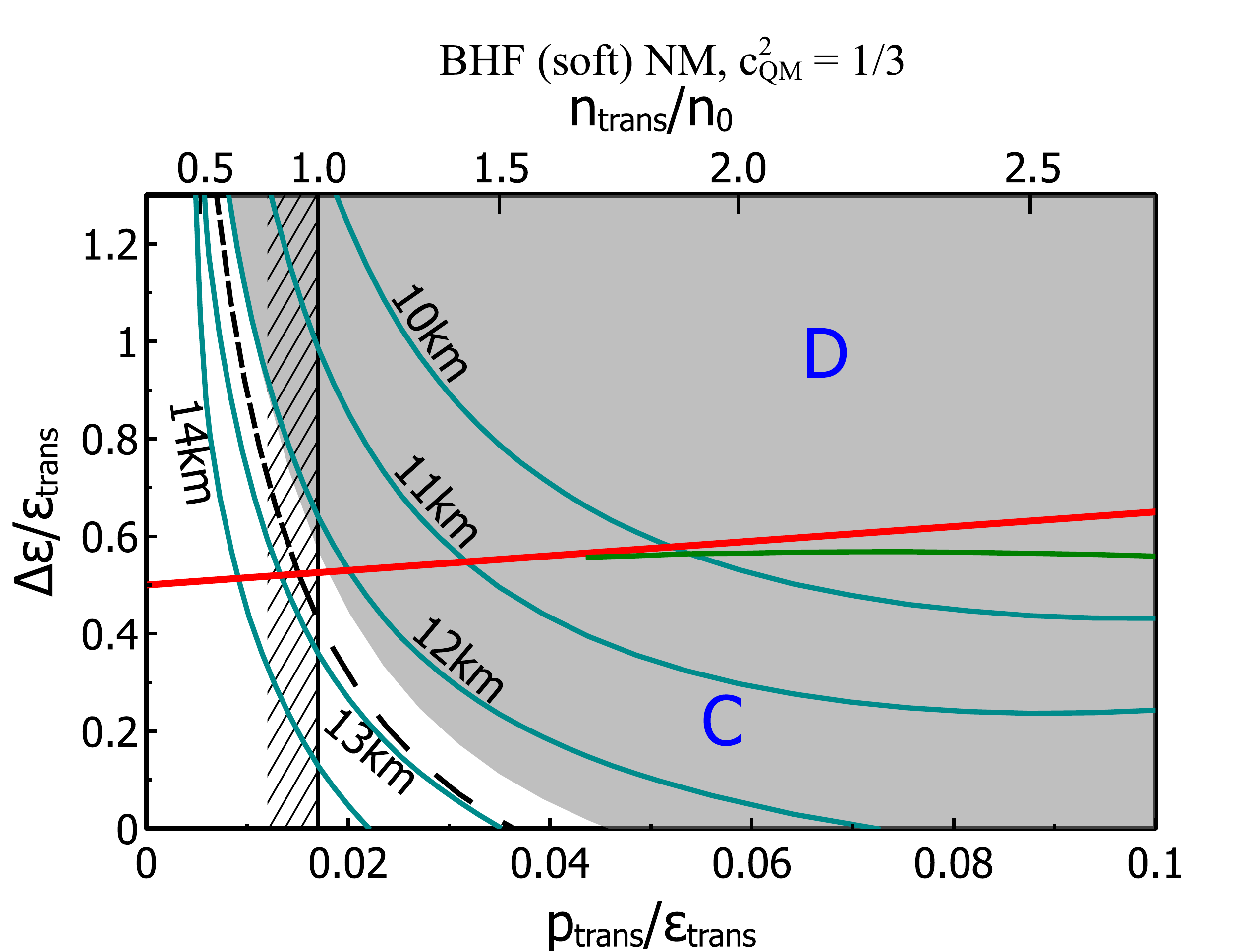}
}
\caption{(Color online). Magnified version of the $\cQMsq=1/3$ plots
in Figs.~\ref{fig:CSS-radius-max-mass}\textendash\ref{fig:CSS-radius}.
In the two left panels, the contours are for the radius of the maximum-mass star, which is typically the smallest star
for the given EoS. In the two right panels, the contours are for
$R_{1.4}$, the radius of a $1.4\,\Msolar$ star. 
The region under and to the left of the hatched bar
is probably unphysical because $\ntrans < n_0$, and it was 
excluded (hatched band) in earlier figures.
The grey shaded region is excluded by the 
observational constraint $\Mmax>2\,\Msolar$. 
The dashed line shows how that region would grow if a
$2.1\,\Msolar$ star were observed.
}
\label{fig:CSS-radius-zoom}
\end{figure*}

For $\cQMsq=1/3$ the allowed region at low transition pressure is small,
so in the right panels of Fig.~\ref{fig:CSS-radius-zoom} we show a magnification
of this region. We see that in the allowed ($\Mmax>2\,\Msolar$ and $\ntrans>n_0$) region 
there is a minimum radius $12.2\,{\rm km}$ for the BHF (soft) hadronic EoS,
and about $12.5\,{\rm km}$ for the DBHF (stiff) hadronic EoS. This
minimum is attained at the lowest possible transition density,
$\ntrans\approx n_0$. As the transition density rises to values
around $2\,n_0$, the minimum radius rises to 12.5 (BHF)
%\com{SH: stops at around $1.74\,n_0$ ($\De \ep = 0$) with $R_{1.4}\approx 12.55 \rm km$} 
or 13.3\,km (DBHF). This is comparable to
the minimum
radius of about 13\,km found in Ref.~\cite{Bedaque:2014sqa}, which
explored a wider range of hadronic EoSs but assumed $\ntrans=2\,n_0$.
These results are consistent with
the lower bound on $R_{1.4}$ for $\cQMsq=1/3$ of about 
11\,km established in Ref.~\cite{Lattimer:2012nd} (Fig.~5)
using the EoS that yields maximally compact stars
(corresponding to CSS with $\ptrans=0$ and $\cQMsq=1/3$) 
obeying  $\Mmax>2\,\Msolar$.
If a $1.4\,\Msolar$
star were observed to have radius below the minimum value, one would
have to conclude that either it is not a hybrid star or that $\cQMsq > 1/3$.

The dashed line shows how the excluded region would grow if a
star of mass $2.1\,\Msolar$ were to be observed. 
This would increase the minimum
radius to about 12.7 (BHF) or 13\,km (DBHF). 

%===============================================================================
\section {The BHF and DBHF EoS of Nuclear Matter}
\label{sec:BHF}

We now discuss in more detail the nuclear matter equations of state that
we use in this work.
We adopt the BHF scheme, in which the only input 
needed is the realistic free nucleon-nucleon (NN) interaction $V$ in the 
Brueckner-Bethe--Goldstone (BBG) equation for the reaction matrix $G$,
\begin{equation}
G[\rho;\omega] = V  + \sum_{k_a k_b} V {{|k_a k_b\rangle  Q  \langle k_a k_b|}
  \over {\omega - e(k_a) - e(k_b) }} G[\rho;\omega], 
  \label{eq:G}
\end{equation}                                                           
\noindent
where $\rho$  is the nucleon number density, and $\omega$ the  starting energy. 
The propagation of intermediate baryon pairs is determined by
the single-particle energy $e(k;\rho) = {{k^2}\over {2m}} + U(k;\rho)$, and the Pauli operator $Q$. Because of the occurrence of $G$ in the single-particle potential
$U(k;\rho) = {\rm Re} \sum _{k'\leq k_F} \langle k k'|G[\rho; e(k)+e(k')]|k k'\rangle_a$,
where the subscript ``{\it a}'' indicates antisymmetrization of the 
matrix element, the BBG equation (Eq.~(\ref{eq:G})) has to be solved in a self-consistent manner
for several  momenta of the particles involved, at the considered densities.

In the nonrelativistic BHF approximation the energy per nucleon is given by \cite{Baldo:1999}
\begin{equation}
{E \over{A}}  =  
          {{3}\over{5}}{{k_F^2}\over {2m}}  + {{1}\over{2\rho}}  
~ \sum_{k,k'\leq k_F} \langle k k'|G[\rho; e(k)+e(k')]|k k'\rangle_a. 
\end{equation}
\noindent
The nuclear EoS can be calculated with good accuracy in the Brueckner two hole-line 
approximation with the continuous choice for the single-particle
potential, and the results in this scheme are quite close to the 
calculations which include also the three hole-line
contribution \cite{Day:1981zz,Song:1998zz,Baldo:2001mv}.  The dependence on the NN interaction, also within
other many-body approaches, has been systematically investigated in Ref.~\cite{Baldo:2012nh}. 

It is well known that,  in order to reproduce the correct saturation point of symmetric nuclear matter, we must
introduce nuclear three-body forces (TBFs). In our approach the TBF is reduced to a 
density-dependent two-body force by averaging over the position of the third particle, assuming that the
probability of having two particles at a given distance is reduced 
according to the two-body correlation function \cite{Baldo:1997ag,Zhou:2004br}.

In this work we use  the Argonne $\rm v_{18}$ NN potential \cite{Wiringa:1994wb},  and the so-called Urbana model for TBFs,  
which consists of an attractive term due to two-pion exchange
with excitation of an intermediate $\Delta$ resonance, and a repulsive phenomenological central term \cite{Carlson:1983kq,Schiavilla:1985gb}. 
Those TBFs produce a shift of about $+1$ MeV in energy and $-0.01$ fm $^{-3}$ in density. This adjustment is obtained 
by tuning the two parameters contained in the TBFs,  and was performed to get an optimal saturation point \cite{Baldo:1997ag,Zhou:2004br}.
At present the theoretical status of microscopically derived TBFs is still quite rudimentary; however a tentative approach has 
been proposed using the same meson-exchange parameters as the 
underlying NN potential \cite{Li:2008bp,Li:2012zzq}.  

Along with the  nonrelativistic BHF EoS we consider its relativistic counterpart, the DBHF scheme \cite{GrossBoelting:1998jg}
where the Bonn-A potential is used for the nucleon-nucleon interaction.
In the low density region ($\rho \rm < 0.3~ fm^{-3}$), the BHF (including TBF) and DBHF equations of state are very similar, 
whereas at higher densities the DBHF is slightly stiffer. 
The discrepancy between the nonrelativistic and relativistic calculation can be easily understood by recalling that the DBHF 
treatment is equivalent  to introducing in the nonrelativistic BHF  the TBF corresponding to the excitation of a nucleon-antinucleon pair, 
the so-called Z-diagram \cite{Brown:1985gt}, which is repulsive at all densities. 
In the BHF treatment with Urbana TBF, both attractive and repulsive TBF are introduced
and therefore a softer EoS is expected.
We report in Table \ref{tab:EoS}  the main properties of both EoSs.

In this work we perform all calculations for
both the BHF and DBHF equations of state for hadronic matter.
This provides a reasonable range of
possible hadronic EoSs, and allows us to
gauge the sensitivity of our results to this source of uncertainty,
although even DBHF is not as stiff as an ultrastiff hadronic EoS such as
DD2-EV \cite{Benic:2014jia} (see Sec.~\ref{sec:maxmass}).

We do not include the effects of hyperons because these are
unknown, and including them would not increase the physical accuracy of our
results. Calculating a hyperonic EoS
requires knowledge of hyperon interactions with other baryons, and
there is little data on hyperon-nucleon interactions \cite{Maessen:1989sx}
and none on hyperon-hyperon interactions or three-body interactions. 
There have been various conjectures about the hyperon interaction
\cite{Schulze2011, Yamamoto2014, Nagels2015, Vidana2011}
and how to include it in BHF \cite{Baldo:1998hd,Baldo:1999rq,Yamamoto2014}
and DBHF \cite{Kata2014} but there is no consensus on the correct result.

\section{Quark Matter via the Field Correlator Method}
\label{sec:FCM}

\subsection{The FCM EoS}
The approach based on the FCM provides a natural treatment
of the dynamics of confinement in terms of the
color electric   ($D^E$ and $D_1^E$)  and color magnetic ($D^H$ and $D_1^H$) 
Gaussian correlators, the former being directly related to confinement,
so that its vanishing above the critical temperature implies deconfinement \cite{DiGiacomo:2000va}.
The extension of the FCM to finite temperature $T$ 
at chemical potential $\mu_q=0$ 
gives  analytical results in reasonable agreement with lattice data,
giving us some confidence that it correctly describes
the deconfinement phase transition
\cite{Simonov:2007xc,Simonov:2007jb}.  In order to derive an EoS of the quark-gluon 
matter in the range of baryon density typical of the neutron star interiors, we 
have to extend  the FCM to nonzero
chemical potential \cite{Simonov:2007xc,Simonov:2007jb}. In this case, the quark pressure for a single 
flavor is simply given by 

\begin{equation}\label{pquark}
P_q/T^4 = \frac{1}{\pi^2} \left[ \phi_\nu \left( \frac{\mu_q - V_1/2}{T} \right) +
\phi_\nu \left(-\frac{\mu_q + V_1/2} {T} \right ) \right ]
\end{equation}
where 
\be
\phi_\nu (a) = \int_0^\infty du  \left({u^4}/{\sqrt{u^2+\nu^2}} \right)
\left(\exp{ \left[ \sqrt{u^2 +\nu^2} - a \right]} + 1\right)^{-1} 
\ee
with   $\nu=m_q/T$,  and $V_1$ is the large distance static $\overline qq$ potential 
whose value at zero chemical potential and temperature is 
$ V_1(T\!=\!\mu_B\!=\!0) = 0.8$ to $0.9$\,GeV 
\cite{Bombaci:2012rv,Plumari:2013ira}.
The gluon contribution to the  pressure is 
\be
\label{pglue}
P_g/T^4 = \frac{8}{3 \pi^2} \int_0^\infty  d\chi \chi^3
\frac{1}{\exp{\left(\chi + \frac{9 V_1}{8T} \right)} - 1}
\ee
and the total pressure is 
\be
\label{pqgp1}
P_{qg} = \sum_{j=u,d,s} P^j_{q} + P_g - \frac{(11-\frac{2}{3}N_f)}{32} \frac{G_2}{2}
\ee
where $P^j_{q}$  and $P_g$ are given in Eqs.~(\ref{pquark}) and  (\ref{pglue}), and  $N_f$ is the number of flavors.
The last term in Eq.~(\ref{pqgp1})
corresponds to the difference of the vacuum energy density in the two phases, $G_2$ being the gluon condensate whose numerical value, determined by the QCD sum rules 
at zero temperature and chemical potential, is known with large uncertainty, $G_2=0.012\pm 0.006~ \rm{GeV^4}$. 
At finite temperature and vanishing baryon density, a comparison with the recent available lattice calculations 
provides clear indications  about the specific values of these two parameters, and in particular 
their values at the critical temperature $T_c$. Some lattice simulations suggest no dependence of $V_1$ on 
$\mu_B$, at least for very small $\mu_B$, 
while different analyses suggest a linear decreasing of $G_2$ with the baryon density $\rho_B$ \cite{Baldo:2003id},
in nuclear matter. However, for simplicity,  in the following  we treat both $V_1$ and $G_2$ as numerical parameters with no dependence on $\mu_B$.

\subsection{The FCM EoS and the CSS parametrization}

\begin{figure}[htb]
\centering
\includegraphics[width=1.15\hsize]{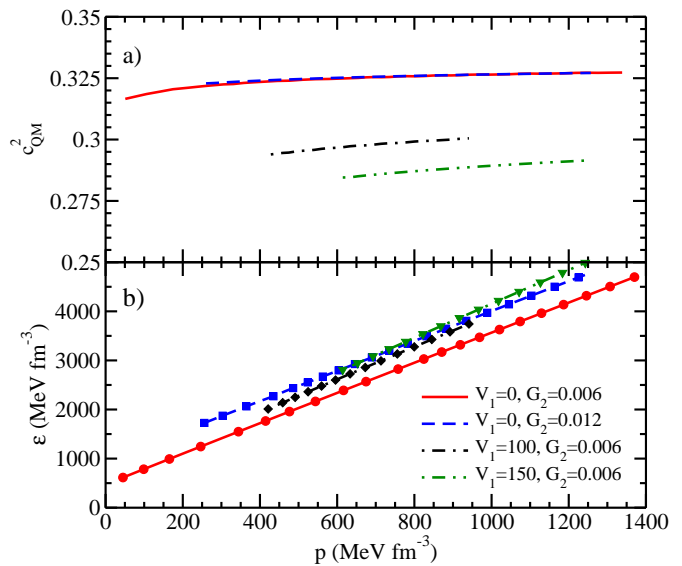}
\caption{(Color online). 
 The squared speed of sound $\cQMsq$ [panel (a)] is displayed vs quark matter pressure for several 
 values  of $V_1$ (in MeV) and $G_2$ (in GeV$^{4}$). In panel (b), the FCM energy density is 
 represented by full symbols, whereas the full lines denote the CSS parametrization 
 given by  Eq.~(\ref{eqn:CSS_EoS}).}
\label{fig:FCM_csq}   
\end{figure}

The CSS parametrization will be applicable to the FCM EoS if the speed of
sound in the FCM EoS depends only weakly on the density or pressure.
In Fig.~\ref{fig:FCM_csq} we show that this is indeed the case.
The upper panel shows the speed of sound vs.~pressure
in the FCM quark matter EoS for different values of the FCM parameters,
displayed in the lower panel. We see that the speed of sound varies by
less than 5\%  over the considered range of pressures along each curve,
and lies in the interval  $0.28 < \cQMsq < 1/3$.  The value of $\cQMsq$ 
shows a weak dependence on  $V_1$ and extremely weak dependence on $G_2$,
which appears as an additive constant in the quark matter EoS according to Eq.~(\ref{pqgp1}).
The transition pressure is more sensitive to the FCM parameters,
increasing rapidly with $V_1$ and with $G_2$. The
energy density at a given pressure increases slightly with 
an increase in $V_1$ or $G_2$.

\begin{figure*}
%\begin{minipage}{15pc}
\includegraphics[width=1.1\hsize]{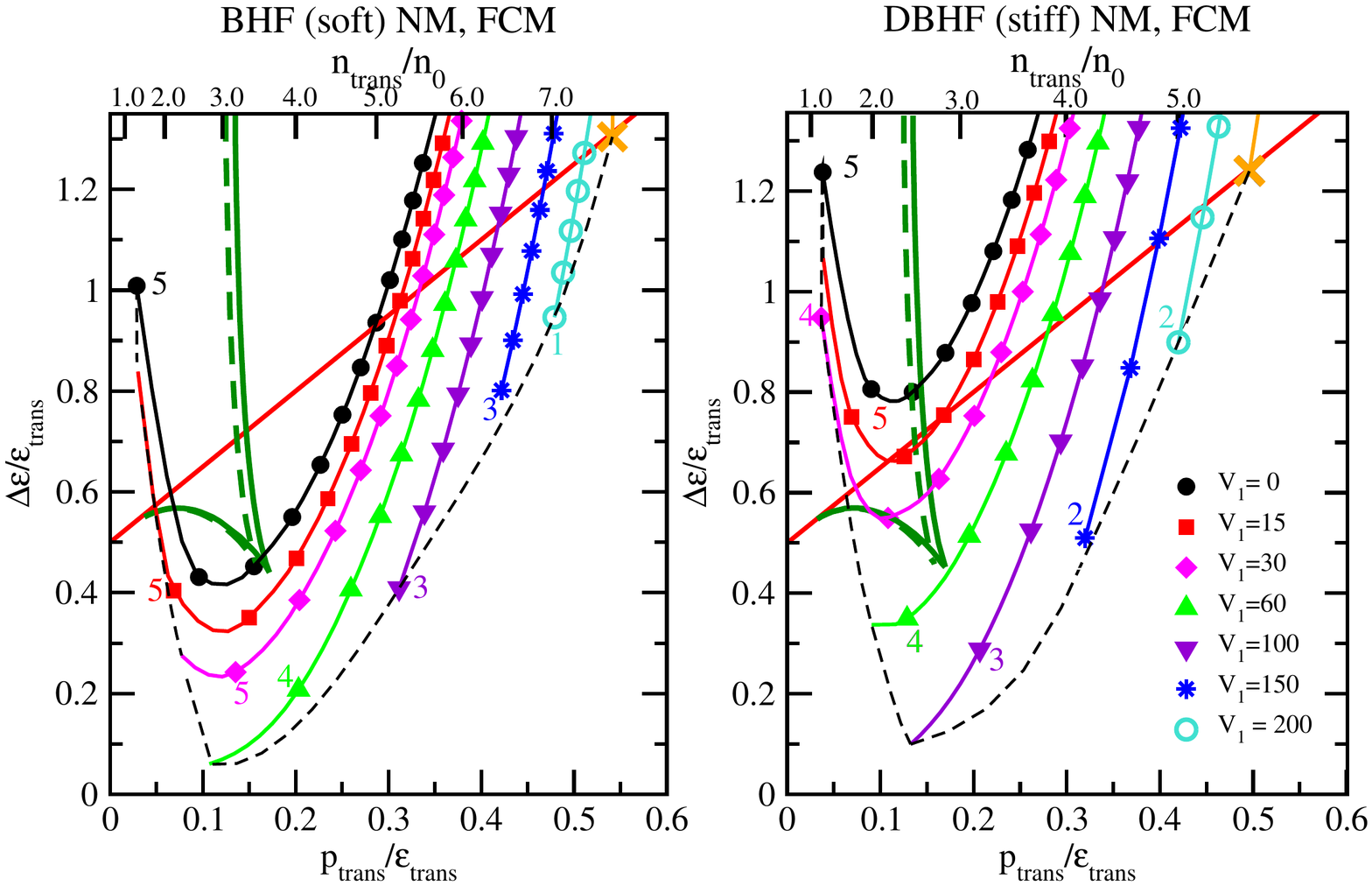}
%\end{minipage} \hspace{5pc}
%\begin{minipage}{15pc}
%\includegraphics[width=8cm]{F4_b.eps}
%\end{minipage} 

\caption{(Color online). The mapping of the FCM quark matter model onto the CSS
parametrization.  Results are obtained using the BHF (left panel) and DBHF
(right panel) nuclear matter EoS. The undecorated curves are the
phase boundaries for the occurrence of
connected and disconnected hybrid branches
(compare Figs.~\ref{fig:phase-diag-schematic} and
\ref{fig:CSS-max-mass}). The thin dashed (black) line and the solid
(black) line studded with circles delimit the region yielded by the
FCM model. Within that region, lines decorated with
symbols give CSS parameter values for
FCM quark matter as $G_2$ is varied at constant $V_1$ (given in MeV).
The (orange) cross denotes the EoS with the  highest 
$\ptrans$, which gives the heaviest FCM hybrid star.
See the text for details. 
}
\label{fig:FCM_map}   
\end{figure*}

\begin{figure*}
\includegraphics[width=\hsize,angle=0]{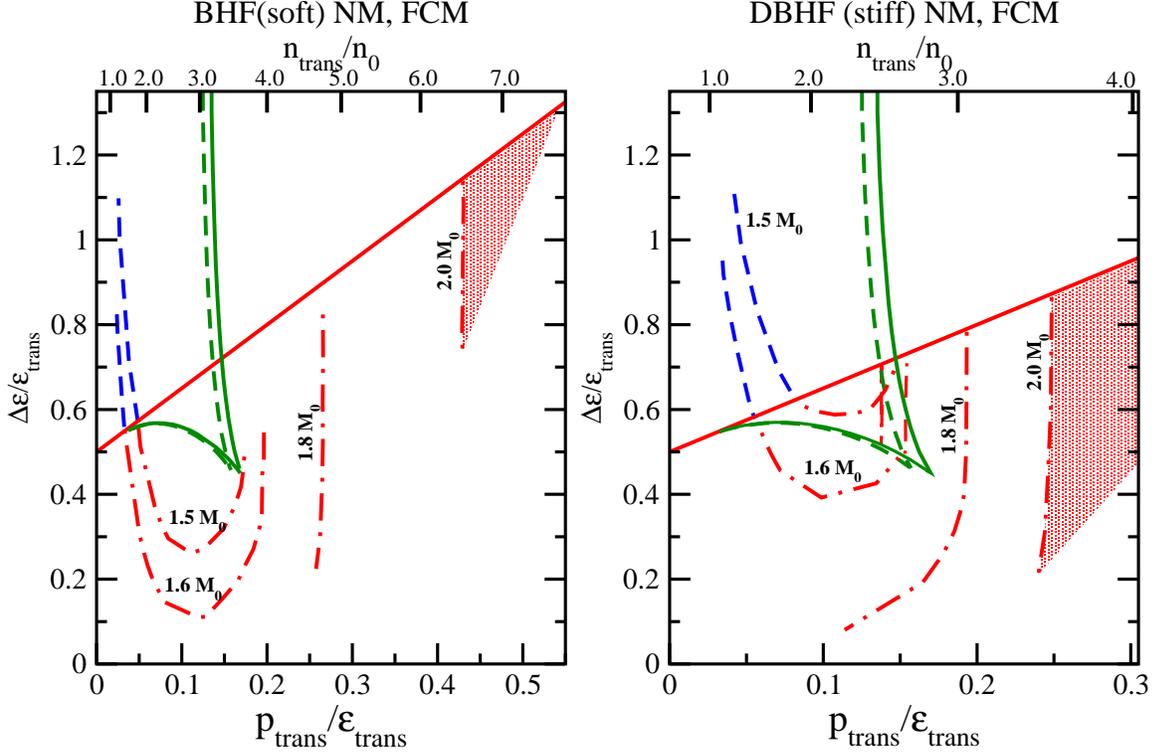}
\caption{(Color online). Contour plots, analogous to Fig.~\ref{fig:CSS-max-mass},
showing the maximum mass of hybrid stars with FCM quark matter
cores, given in terms of the corresponding CSS parameter values rather than
the original FCM parameter values. As in Fig.~\ref{fig:FCM_map},
solid lines are phase boundaries (compare Figs.~\ref{fig:phase-diag-schematic} and \ref{fig:CSS-max-mass}).
The shaded sectors indicate the parameter regions accessible by the FCM and with $M_{\rm max}>2\,\Msolar$.
In each panel the lower border of the shaded region meets the phase boundary (red line) at the point with highest value of $V_1$ 
reported in Fig.~\ref{fig:FCM_map} as an orange cross.
Note the different scales on the x axis for the two panels.
}
\label{fig:FCM-max-mass}   
\end{figure*}

\begin{figure}
\centering
\includegraphics[width=1.4\hsize, angle=0]{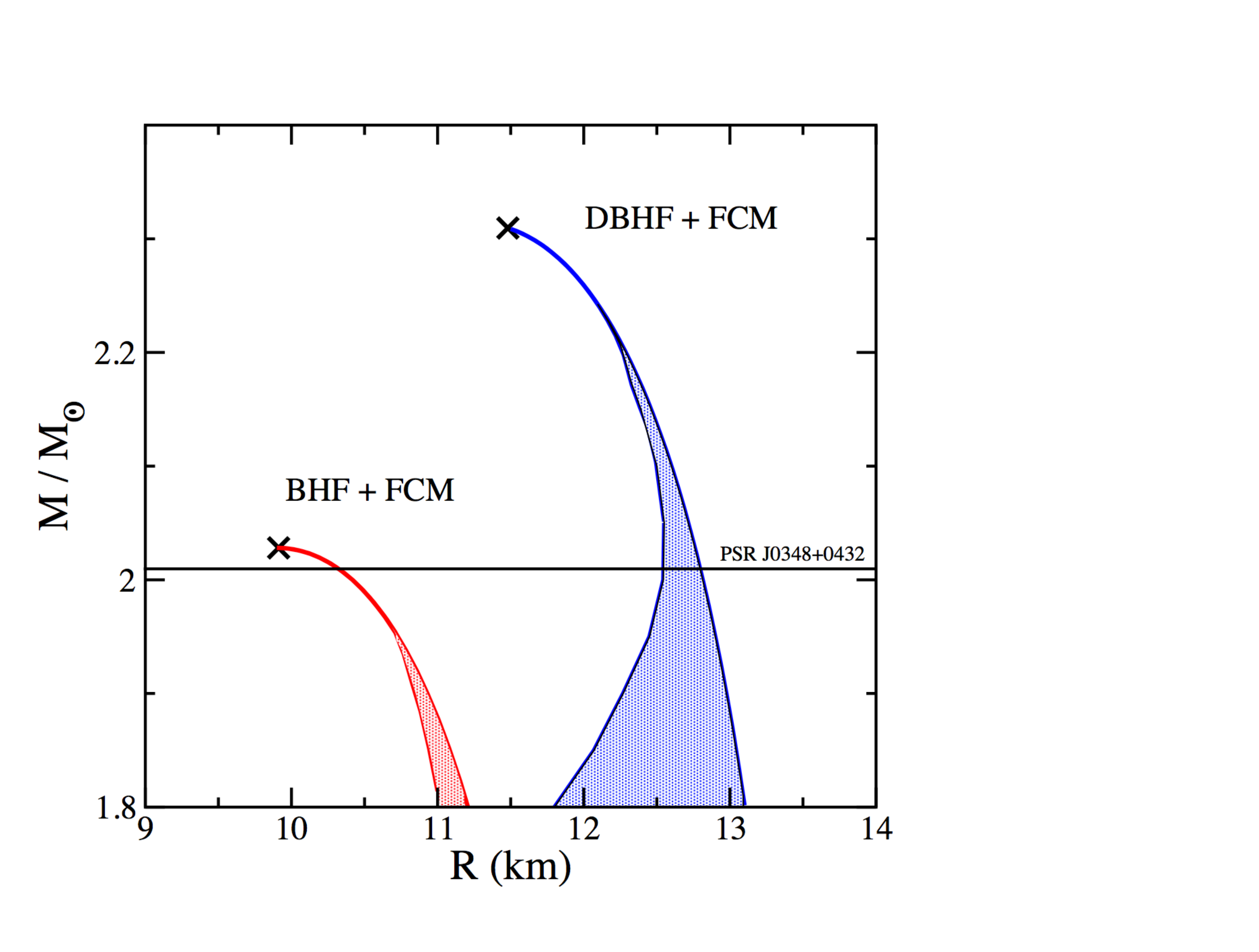}
\caption{(Color online). Shaded areas show range of radii of stars with a given maximum mass
when varying FCM parameters.
The thick dashed lines indicate the purely hadronic mass-radius configurations. The crosses at the top of the shaded regions correspond to the maximal configuration 
indicated by the same symbol in Fig.~\ref{fig:FCM_map}. 
The observational constraint \cite{Antoniadis:2013en}  on the star mass is indicated by a horizontal line.
}
\label{fig:MR-regions}   
\end{figure}

To illustrate how well the CSS parametrization fits the FCM EoS, we show in
the lower panel of Fig.~\ref{fig:FCM_csq} that, for the same FCM parameter choices,
we can always find suitable values of the CSS parameters which fit the FCM calculation
extremely well. This means that there exists a well-defined
mapping between the FCM parameters ($V_1$, $G_2$) and the CSS parameters
($\ptrans/\etrans$, $\De\ep/\etrans$, $\cQMsq$). Note that
the mapping depends on the EoS of the hadronic matter.

The mapping is displayed in Fig.~\ref{fig:FCM_map}, which shows the region of
the CSS parameter space where FCM equations of state are found. 
As in the phase diagrams in Sec.~\ref{sec:CSS-constraints},
we show the plane  whose coordinates are the CSS parameters
$\De\ep/\etrans$ and $\ptrans/\etrans$. For the hadronic EoS we use
BHF (left panel) and DBHF (right panel).  The lines without points
represent the phase boundaries, as for the figures in
Sec.~\ref{sec:CSS-constraints}, for connected and disconnected branches.
Whether a given FCM EoS yields stable hybrid
stars depends on which of those phase regions
(see Fig.~\ref{fig:phase-diag-schematic}) it is in.
The solid (green) phase boundary with a cusp at $\ptrans/\etrans 
\approx 0.17$ delimits the region
with a disconnected branch for $\cQMsq=1/3$,
while the nearby dashed (green) line is for
$\cQMsq=0.28$, so these  span the
range of $\cQMsq$ relevant for the FCM, as discussed in Fig.~\ref{fig:FCM_csq}.  
It is evident that the dependence on $\cQMsq$ is tiny and negligible for
practical purposes.  

The thin dashed (black) line and the solid (black) line studded with 
circles delimit the equations of state yielded by the FCM calculation. 
Within that region,
the lines studded with points show the CSS parametrization
of the FCM quark matter EoS, where along each line we keep $V_1$ 
constant and vary $G_2$. Above that region, which corresponds to negative values of $V_1$, the EoS cannot reproduce the $2\,\Msolar$ limit, and in this sense is unphysical (Fig.~\ref{fig:FCM-max-mass}). Below that region, there would be no transition from hadronic to quark matter, as explained below.

In Fig.~\ref{fig:FCM_map}, $V_1$ varies from 0 up to the maximum value at which hybrid star configurations occur, which is indicated by an (orange) cross.
For the BHF case that value is
$V_1=240\,\,\MeV$, $G_2=0.0024\,\,\GeV^{4}$ and for the  DBHF case 
it is $V_1=255\,\,\MeV$, $G_2=0.0019\,\,\GeV^{4}$.
Along each FCM curve in Fig.~\ref{fig:FCM_map}
the parameter  $G_2$ starts at the minimum
value at which there is a phase transition from hadronic to FCM 
quark matter; at lower  $G_2$ the quark and the hadronic pressures  $p(\mu)$ do not cross at any $\mu$. 
On each curve one point is labeled with its value of $G_2/(10^{-3}\,{\rm GeV}^4)$, and subsequent points are at intervals where $G_2$ increases
in increments of 1 in the same units.

We observe that along each line of constant $V_1$, $\ptrans/\etrans$ grows with $G_2$. 
This can be explained by recalling the linear dependence of the quark pressure on $G_2$ in Eq.~(\ref{pqgp1}), so that, at fixed chemical potential, 
an increase of $G_2$ lowers the quark pressure, making quark matter less favorable,
and shifting the transition point to higher chemical potential or pressure.
This was already discussed in Ref.~\cite{Baldo:2008en} for 
BHF nuclear matter, and is equally applicable to DBHF nuclear matter.
Obviously if $G_2$ becomes too small, the phase transition takes place in a region of low 
densities where finite nuclei are present, and the homogeneous nuclear matter approach 
becomes invalid. 

The qualitative behavior of the curves of constant $V_1$
can be understood in terms of the Maxwell construction between the purely hadronic phase and the quark phase. 
The fact that $\Delta \epsilon/\etrans$ goes through a minimum
(which is always at $\ptrans/\etrans\approx 0.1$) as $G_2$
is increased at constant $V_1$ 
can be understood from  Fig.~2 of Ref.~\cite{Baldo:2008en},
which shows pressure $p$ as a function of baryon density $n$ 
and the location of the hadron (BHF EoS)
to quark (FCM EoS) transition when $G_2$ is varied.
The hadronic EoS is strongly curved, especially at low pressure,
while the FCM EoS is closer to a straight line. Consequently,
the baryon density difference between the two phases
at a given pressure has a minimum at densities around $2\,n_0$,
which corresponds to  $\ptrans/\etrans\approx 0.1$.
As $G_2$ increases, the transition pressure rises, scanning
through this minimum. It follows that
the energy density difference also goes through a minimum, because
$\epsilon=\mu n - p$, and $p$ and $\mu$ are continuous at the transition,
so $\De\ep = \mu\De n$.
The DBHF hadronic EoS is very similar to BHF at
low pressure, so the curves have their minima at the same value of
$\ptrans/\etrans$ in both panels of Fig.~\ref{fig:FCM_map}.

We also see in Fig.~\ref{fig:FCM_map} that an increase of $V_1$ moves the
curves slightly downward and to the right.  This is expected since $V_1$ is a
measure of the interparticle strength, and is
inversely proportional to the pressure of the system, so that the pressure
decreases as $V_1$ is increased at fixed $\mu$, and, as already discussed for
the parameter $G_2$, a decrease of the quark pressure raises $\ptrans$.  The
role of $V_1$ and $G_2$ in the quark EoS discussed so far, provides in the
same way a qualitative understanding of $\cQMsq$ in panel (a) in
Fig.~\ref{fig:FCM_csq}, although, as already noticed, the effect in
Fig.~\ref{fig:FCM_map} of the change in $\cQMsq$ is negligible.

\subsection{Expected properties of mass-radius curves}

By comparing Fig.~\ref{fig:FCM_map} with Fig.~\ref{fig:phase-diag-schematic}
we can see that when combining FCM quark matter with
BHF (soft) nuclear matter, the physically allowed range of
FCM parameter values yields EoSs that are mostly in
regions C and A, where there is no
disconnected hybrid branch. At the lowest transition densities the FCM EoS can
achieve a large enough energy density discontinuity to yield a disconnected
branch (region D).

For the DBHF (stiff) nuclear EoS there is a wider range of values of
$V_1$ and $G_2$ that give disconnected branches, and some of them 
give simultaneous connected and disconnected branches.
This difference can be understood in terms of the stiffness of the EoSs.
A change from a soft hadronic EoS  (BHF)
to a stiff one (DBHF) produces a steeper growth of the hadronic pressure
as a function of the baryon density. Referring again to
Fig.~2 of Ref.~\cite{Baldo:2008en}, this pulls the DBHF
$p(n)$ curve further away from the FCM curve, giving a
larger difference in baryon density at a given pressure, and hence, 
as noted above, a larger $\De\ep$. This is why the curves for
DBHF+FCM (right panel of Fig.~\ref{fig:FCM_map}) are shifted upwards along the
 $\De\ep/\etrans$ axis compared to the BHF+FCM curves 
(left panel of Fig.~\ref{fig:FCM_map}).

We can
calculate the maximum mass of a hybrid star containing a FCM core
as a function of the FCM parameters, and then use the mapping described
above to obtain the CSS parameter values for each FCM EoS,
producing a contour plot of maximum mass (Fig.~\ref{fig:FCM-max-mass}) 
for BHF (left panel)
and DBHF (right panel) hadronic EoS. Given that the CSS parametrization
is a fairly accurate representation of the FCM EoS, one would expect this
to be very similar to the corresponding plot for CSS itself with
$\cQMsq=1/3$ (Fig.~\ref{fig:CSS-max-mass}), and this is indeed the case.
The contours in Fig.~\ref{fig:FCM-max-mass} are restricted to the region
corresponding to physically allowed FCM parameter values, so they end at the edges of that region.

The triangular shaded area at the edge of each panel
shows the region 
of the parameter space that is accessible by the FCM and 
is consistent with the measurement of a $2\,\Msolar$,
by having hybrid stars of maximum mass greater than $2\,\Msolar$.
The (orange) cross in each panel of Fig.~\ref{fig:FCM_map} is at the 
high-transition-pressure corner of that triangular area.
The heaviest BHF+FCM hybrid star has a mass of
$2.03\,\Msolar$, and the heaviest DBHF+FCM hybrid star has a mass of
$2.31\,\Msolar$.

As noted in Sec.~\ref{sec:maxmass}, the hybrid stars in this
physically allowed and FCM-compatible region
of the phase diagram lie on a very tiny connected branch, 
covering a very small range of central pressures and masses and radii,
and would therefore occur only rarely in nature. These stars
have very small quark matter cores
(see Ref.~\cite{Alford:2013aca}, Figs.\,5\textendash 6), and their mass and
radius are very similar to those of the heaviest purely hadronic star,
but there could be other clear signatures of the presence of the
quark matter core, such as different cooling behavior.

The CSS parametrization has another region where heavy hybrid stars
occur, at low transition pressure (see Fig.~\ref{fig:CSS-max-mass}),
but the FCM does not predict that the quark matter EoS could be
in that region.

 To characterize the radius of FCM hybrid stars we cannot construct
contour plots like Fig.~\ref{fig:CSS-radius} because, as we have just seen,
the FCM predicts that only hybrid stars with mass very close to the maximum
mass are allowed. There are no FCM hybrid stars with mass around 
$1.4\,\Msolar$.  Instead, in Fig.~\ref{fig:MR-regions} we
show the range of radii of stars with a given maximum mass
when varying FCM parameters, for our two different hadronic EoSs.
The right-hand edge of each shaded region traces out
the mass-radius relation for hadronic stars with the corresponding
hadronic EoS.
The FCM hybrid stars form very small connected branches
which connect to the nuclear matter where the central
pressure reaches the transition pressure (see Sec.~\ref{sec:maxmass}),
so the hybrid stars do not deviate very far from the hadronic mass-radius
curve. Hence the shaded regions in Fig.~\ref{fig:MR-regions} are narrow,
especially in the observationally allowed ($M_{\rm max}>2\,\Msolar$) region, which perfectly matches the prediction of CSS parametrization on the maximum-mass star radius in the high-transition-pressure region (see the left panels of Fig.~\ref{fig:CSS-radius-max-mass}). For BHF (soft) nuclear matter, the hadronic stars, and hence the hybrid stars,
are smaller because the nuclear mantle is more compressed by the self gravitation of the star.

\section{Conclusions}
\label{sec:conclusions}

We have shown how observational constraints on the mass and 
radius of hybrid stars can be expressed as constraints on the
parameters of the CSS parametrization of the high-density EoS,
which, in the space of possible models of quark matter, is a
reasonably model-independent parametrization.
Of course, physical predictions from CSS 
depend on the hadronic EoS with which it is combined.
The CSS parametrization
assumes a sharp transition from nuclear matter to a
high-density phase such as quark matter, and that the speed of sound
in that phase is independent of the pressure.
We found that the observation of a $2\,\Msolar$ star
constrains the CSS parameters significantly 
\cite{Alford:2013aca,Bedaque:2014sqa}.

If, as predicted by many physical models of quark matter,
$\cQMsq\lesssim 1/3$, then for typical models of hadronic matter
such as BHF or DBHF there are two possible scenarios
(we discuss ultrastiff hadronic EoSs below).

First, there is a low-transition-pressure scenario, where
the transition to the high-density phase occurs at
$\ntrans \lesssim 2\,n_0$
(the unshaded region on the left side of the two left panels of 
Figs.~\ref{fig:CSS-max-mass}\textendash\ref{fig:CSS-radius}). In this scenario,
the hybrid branch of the mass-radius relation will be connected to the 
nuclear branch.
In the $\cQMsq\lesssim 1/3$ and low-transition-pressure scenario 
there are strong
constraints on the radius of the star, as shown in
Fig.~\ref{fig:CSS-radius-zoom}. The radius of the maximum-mass
star (which is typically the smallest possible star) must be greater
than about 11.5\,km, and the radius of a $1.4\,\Msolar$
star must be greater than about 12.2\,km \cite{Lattimer:2012nd}.
For a stiffer hadronic EoS, these minima are raised by about 0.15 to 0.3\,km. If a neutron star of mass $2.1\,\Msolar$ were observed then this
constraint would tighten, increasing the minimum radius to about
12.1\,km.
If a star smaller than the minimum radius were observed, we
would have to conclude that either the transition is
outside the low-density regime or $\cQMsq>1/3$.
Conversely, if theoretical considerations established
that $\cQMsq$ is smaller than $1/3$, the
minimum radius would become larger \cite{Bedaque:2014sqa}.

Secondly, there is a high-transition-pressure scenario (the white region on the
right side of the left panels of Figs.~\ref{fig:CSS-max-mass}\textendash\ref{fig:CSS-radius}).
This tends to give a very small branch of hybrid stars
with tiny quark matter cores,
occurring in a narrow range of central pressures just above the
transition pressure. This is why the mass and radius contours
become almost vertical in this region: the hybrid star has almost the same
mass and radius as the heaviest purely hadronic star (the one where
the central pressure is $\ptrans$), and so the properties of these hybrid stars
depend on the hadronic EoS (see Fig.~\ref{fig:MR-regions}) via
$\ptrans/\etrans$ but not on quark matter properties such as
$\De\ep$ or $\cQMsq$.

If the hadronic matter is extremely stiff (e.g. DD2-EV 
\cite{Benic:2014jia})
or the quark matter has
$\cQMsq$ larger than $1/3$ then a larger region of the CSS parameter
space becomes allowed. The right panels of Figs.~\ref{fig:CSS-max-mass}\textendash\ref{fig:CSS-radius} show the extreme case where 
$\cQMsq=1$. In this case the minimum possible radius is 9.0\,km.

Disconnected hybrid branches are of special interest, because they
give a characteristic signature in mass-radius measurements.
For both the hadronic EoSs that we study,
disconnected hybrid star branches are excluded by the $\Mmax>2\Msolar$
constraint
for $\cQMsq\leqslant 1/3$, and even for
larger $\cQMsq$ they only arise if the hadronic matter EoS is
stiff. Explorations of the ultrastiff hadronic  DD2-EV EoS
indicate that disconnected hybrid star branches can occur at
moderate $\cQMsq$, and it would be interesting to include this EoS
in a future study.

Our work is intended to 
motivate the use of the CSS parametrization as a framework 
in which the implications of observations of neutron stars
for the high-density EoS can be expressed and discussed in a way
that is reasonably independent of the modeling of the EoS of the
high-density phase (quark matter in our case) 
\cite{Benic:2014jia, Alvarez-Castillo:2014nua}.

As an application to a specific model, we performed calculations for the FCM
quark matter EoS. We showed that the FCM equation of state can be accurately
represented by the CSS parametrization, and we displayed the mapping between
the FCM and CSS parameters. We found that FCM quark matter has a
speed of sound in a narrow range around $\cQMsq=0.3$, and the FCM 
family of EoSs covers a limited region of the space of all possible
EoSs (Fig.~\ref{fig:FCM_map}).
Once the  observational constraint $M_{\rm max}>2\,\Msolar$
is taken into account, the allowed region in the parameter space
is drastically reduced to the shaded areas of Fig.~\ref{fig:FCM-max-mass}.
This corresponds to the high-transition-pressure scenario, with a small 
connected branch of hybrid stars with tiny quark matter cores.
Such stars would be hard to distinguish from hadronic stars via
mass and radius measurements, but the quark matter core could be
detectable via other signatures, such as cooling behavior.
These hybrid stars have central densities
larger than $6.5\,n_0$ in the BHF case, and $3.5\,n_0$ in the DBHF case
(Fig.~\ref{fig:FCM-max-mass}) which means that,
according to Refs.~\cite{Schulze2011, Yamamoto2014, Vidana2011, Kata2014},
hyperons could play an important role in the BHF case,
and they cannot be ruled out even in the relativistic DBHF case.
As discussed in Sec.~\ref{sec:BHF}, we ignored hyperons because
 we are already using two different hadronic EoSs, 
one stiff and one soft, to estimate the sensitivity of our results
to the hadronic EoS, and the effect of hyperons remains unknown.
In the future we hope that more experimental
data will constrain the high-density hadronic EoS, including the
hyperonic contribution.

\section{Acknowledgments}
This material is based upon work supported by the U.S. Department of Energy (DOE), Office of Science, Office of Nuclear Physics under Award No. DE-FG02-05ER41375, and by the DOE Topical Collaboration 
 ``Neutrinos and Nucleosynthesis in Hot and Dense Matter''
Grant No. DE-SC0004955. Partial support comes from ``NewCompStar,'' COST Grant No. MP1304. We thank J. Lattimer, F. Weber, D. Blaschke, and M. Prakash for
helpful discussions. 

% Override the revtex href command in order that the JHEP bib style
% will work properly:
\renewcommand{\href}[2]{#2}

% macros used by ADS Database BiBTeX entries:
% see http://adsabs.harvard.edu/abs_doc/aas_macros.sty
\newcommand{\apjl}{Astrophys. J. Lett.\ }
\newcommand{\mnras}{Mon. Not. R. Astron. Soc.\ }
\newcommand{\aap}{Astron. Astrophys.\ }

\bibliographystyle{JHEP_MGA}
\bibliography{map_fcm} 
 
\end{document}